\documentclass[11pt]{article}

\usepackage{fullpage}
\usepackage{graphicx}
\usepackage{caption,subcaption}
\usepackage{setspace,hyperref}
\usepackage{amsmath,amssymb}
\usepackage{epstopdf}
\usepackage{todonotes}

\newcommand{\mcubed}{\hspace{1mm}\mathrm{m}^{3}}
\newcommand{\Sv}{\hspace{1mm}\mathrm{Sv}}
\newcommand{\mneg}{\hspace{1mm}\mathrm{m}^{-3}}
\newcommand{\degC}{\hspace{1mm}^{\circ}\mathrm{C}}
\newcommand{\kg}{\hspace{1mm}\mathrm{kg}}
\newcommand{\x}{\mathrm{x}}
\newcommand{\Hp}{H_{\mathrm{pert}}}
\newcommand{\Tp}{T_{\mathrm{pert}}}

\begin{document}

\title{Basin bifurcations, oscillatory instability and rate-induced thresholds for AMOC in a global oceanic box model}

\author{Hassan Alkhayuon$^{1}$, Peter Ashwin$^{1}$, Laura C Jackson$^{2}$, Courtney Quinn$^{1,3}$ and Richard A Wood$^{2}$}

\maketitle

\noindent $^{1}$Department of Mathematics, University of Exeter, Exeter EX4 4QF, U.K.\\
$^{2}$Met Office Hadley Centre, Exeter, U.K.\\
$^{3}$CSIRO Oceans and Atmosphere, Hobart, TAS, Australia




\begin{abstract}
The Atlantic Meridional Overturning Circulation (AMOC) transports substantial amounts of heat into the North Atlantic sector, and hence is of very high importance in regional climate projections. The AMOC has been observed to show multi-stability across a range of models of different complexity. The simplest models find a bifurcation associated with the AMOC ``on'' state losing stability that is a saddle node. Here we study a physically derived global oceanic model of Wood {\em et al} with five boxes, that is calibrated to runs of the FAMOUS coupled atmosphere-ocean general circulation model. We find the loss of stability of the ``on'' state is due to a subcritical Hopf for parameters from both pre-industrial and doubled CO${}_2$ atmospheres. This loss of stability via subcritical Hopf bifurcation has important consequences for the behaviour of the basin of attraction close to bifurcation. We consider various time-dependent profiles of freshwater forcing to the system, and find that rate-induced thresholds for tipping can appear, even for perturbations that do not cross the bifurcation. Understanding how such state transitions occur is important in determining allowable safe climate change mitigation pathways to avoid collapse of the AMOC.
\end{abstract}






\section{Introduction}
\label{sec:intro}

Since Stommel's realisation \cite{Stommel1961} that competing thermal and haline circulation effects may give rise to multiple stable states in the Atlantic Meridional Overturning Circulation (AMOC), several studies have found evidence of this across a wide range of scales in the model hierarchy: see for example \cite{Dijkstra2007,Dijkstra2013,hawkins2011,Rahmstorf1996,Rahmstorf1999,rahmstorf2005thermohaline}. Dansgard-Oeschger (DO) events give strong palaeoclimatic evidence of repeated and significant changes to the transport properties of AMOC between an ``on'' state that transports significant amounts of heat to high latitudes and an ``off'' state that does not. In the palaeoclimate record, the time of residence in the states can be a matter of 1000s of years while the switching can be relatively rapid and may be associated with Heinrich events where there were large inputs of fresh water into the Atlantic from glaciers \cite{Bondetal:2013,ClementPeterson2008}.  It is possible that fresh water input at high latitudes (i.e. changes to precipitation and evaporation patterns or ice sheet melting) could give a perturbation that can move the AMOC over a tipping point from ``on'' to ``off'' state. 

Tipping points are significant changes to the solutions of a system in response to a small perturbation, and there has been a concerted effort to understand their appearance and predictability \cite{Scheffer2009,Lenton2011}. They are often thought of in terms of bifurcations for some slowly varying nonlinear system, though \cite{Ashwin2011} point out that, in addition to such bifurcation-induced tipping (B-tipping), noise fluctuations (N-tipping) and sufficiently rapid shifts of a parameter can give rise to similar phenomena, where the tipping is associated with critical rates \cite{Scheffer2008} (R-tipping) rather than specific bifurcations. 

For the climate, a change in external forcing (e.g. atmospheric greenhouse gases or fresh water input from ice sheets) would be expected to change both the current equilibrium state (``on'') of the AMOC and its basin of attraction. In some cases the forcing could be strong enough that the ``on'' state is no longer a valid equilibrium. In that case the AMOC would be expected to collapse to the ``off'' state (B-tipping), although if the forcing were reversed quickly enough it could recover \cite{vdB2018} (with some ``resilience time'' \cite{JW17}). For weaker forcing, the ``on'' state may remain a viable equilibrium. If the forcing changes sufficiently slowly, the AMOC will keep up with the changing equilibrium and remains within its basin of attraction, whereas if the change in forcing is too fast, the current AMOC state may lie outside the basin of the new ``on'' equilibrium. In this latter case the AMOC may be expected to transition to an ``off'' equilibrium, even if the ``on'' state still exists and is stable. Such rate dependent tipping of the AMOC has already been observed in a an intermediate-complexity climate model \cite{stocker1997}: when atmospheric carbon dioxide was increased to a new, steady concentration value slowly, the AMOC remained on, but when it was increased more quickly to the same steady concentration, the AMOC collapsed.

For the AMOC, one hopes that monitoring fluctuations in its strength \cite{mccarthy2015measuring}, or other related ocean variables \cite{RWSH17}, will enable us to spot early warning signals of an approaching tipping point due to changing parameters in the system. These changing parameters are typically idealised as changes in freshwater input or ``hosing'' of regions of the North Atlantic. We note that some more complex models (e.g. \cite{Peltier:2014}) can simulate repeated switching between these AMOC states as slow and spontaneous relaxation oscillations without the need for any freshwater input.

For simple ocean box models such as Stommel's \cite{Stommel1961}, the ocean is divided into a small number of boxes and continuity of salt and heat fluxes is used to derive a number of dynamic equations. For these models it is possible to perform a fairly complete analysis of the dynamics and find steady bifurcation points where changes to model parameters give changes in the qualitative set of solutions, although even for these models there can be surprises, for example in the susceptibility to noise \cite{Lohmann1999}. These have been extended by some authors to include multiple-box models of the oceans \cite{LucariniStone2005,Titzetal2002a} where the increased dimension of phase space means that bifurcations can involve dynamically more complex states. In particular, \cite{Titzetal2002b,Titzetal2002a} find the destabilisation of the ``on'' state may involve Hopf and homoclinic bifurcations. At the other end of the modelling scale, the most realistic models use detailed global atmosphere-ocean general circulation models (AOGCMs) with realistic geometry, and earth system models of intermediate complexity (EMICs) which are similar to AOGCMs but use simplified physical schemes to speed up computation. A few studies have shown the existence of multiple AMOC equilibria and transitions between the states, usually in response to idealised ``hosing'' experiments in which additional fresh water input is applied to the North Atlantic. This is either slowly increased up to and beyond a critical value (e.g. \cite{hawkins2011,jacksonfamous17,rahmstorf2005thermohaline}), or applied as a large perturbation for a limited time (e.g. \cite{DeVries05,JW17,liu2012diagnostic,manabe88}. A few AOGCM studies have begun to explore the transient AMOC dynamics as thresholds are crossed, including the resilience time when fresh water forcing is temporarily increased to a level where the AMOC ``on'' state is unsustainable \cite{jackson18ClDyn,JW17,sijp2012precise,LucariniCalmanti2005,LucariniCalmanti2007}.  

As mathematical models of AMOC across scales may have quite different structures and parametrizations, it can be a challenge to make direct quantitative comparisons between the models at the ends of the scale. Analysis of AOGCM solutions requires a substantial reduction of complexity, for example by considering time- and spatially-averaged flows. In an attempt to make direct quantitative comparison between the most complex and simple models, a simple but physically-based five box model for global ocean circulation that exhibits AMOC switching has been developed by Wood et al. \cite{RWSH17}. They show that the dynamics of the AMOC in a hosing experiment with the FAMOUS AOGCM \cite{Smith2012,hawkins2011,jacksonfamous17} are well described by the dynamics of the box model, when the boxes are calibrated to the water masses in the unperturbed AOGCM state. This allows direct quantitative comparisons between states and fluxes for the AOGCM runs and the much simpler five box model, so that the box model can be used to provide insights into the AOGCM's behaviour.   

In this study we investigate in detail the dynamics of this box model of \cite{RWSH17}, using our analysis to gain insight into possible tipping behaviours of the AMOC and hence to interpret the behaviours seen in more complex AOGCM experiments. We use a timescale separation that further reduces the five box model of \cite{RWSH17} to a three box model that seems to include the qualitative and quantitative behaviour of interest for the five box model. We find notable departures from the dynamics of the two-box model \cite{Stommel1961}, some of which have been found in other multi-box models \cite{LucariniStone2005,Titzetal2002b,Titzetal2002a}. The differences are:
\begin{itemize}
\item The first loss of stability of the on-state is via a subcritical Hopf bifurcation rather than a fold bifurcation. This distance between the subcritical Hopf and fold bifurcations increases on increasing CO$_2$. Although there is a fold nearby, it represents the meeting of two solutions that are both unstable: cf \cite{Titzetal2002b,Titzetal2002a}.
\item There can be an oscillatory relaxation to the on-state found both in three box model and the AOGCM runs: cf \cite{LucariniStone2005}.
\item The basin of attraction of the ``on'' state goes from semi-infinite size to finite size at a homoclinic bifurcation.
\end{itemize}
These changes in the basin of attraction of the ``on'' state mean that the response of the system to changes in forcing can vary with time in a non-trivial manner. We find rate-dependent effects where thresholds involve interaction between the attractors of the system and the unstable states. In climate terms, these insights into the dynamics of the model system can potentially lead to observable indicators of when the AMOC is approaching, or has crossed, critical stability thresholds; for example trajectories of the subtropical and subpolar Atlantic salinity undergo a qualitative change in character when the current, strong AMOC becomes unsustainable. The possibility of rate-dependent tipping has important implications for determining safe emissions pathways to reach climate stabilisation targets. 

The paper is organized as follows. In Section~\ref{sec:concept} we briefly review box models of the AMOC and their bifurcations. Section~\ref{sec:fivebox} introduces the five box model of \cite{RWSH17} and presents an empirical three box reduction to fit numerical observations that two of the boxes show very little in the way of dynamic variation. We perform a comparative bifurcation analysis of these models for parameters from \cite{RWSH17} that are fit to FAMOUS runs with pre-industrial ($1\times $CO$_2$) and double ($2\times $CO$_2$), and find good agreement in both cases.  Section~\ref{sec:tipping} discusses some of the implications for tipping points caused by time dependent changes to the hosing that may lead to collapse of AMOC. We compare these time-dependent perturbations to similar AOGCM studies \cite{JW17,RWSH17}. In particular, Section~\ref{sec:Btip} demonstrates that crossing the threshold need not inevitably lead to tipping: there may be temporary resilience as long as and crossing is reversed sufficiently rapidly, and we quantify how fast this reversal must be. Section~\ref{sec:Rtip} shows that on the other hand, too rapid a change in forcing can destabilise the system, even if the bifurcation is not crossed. In Section~\ref{sec:gcm} we relate the results of our box model analysis to relevant AOGCM experiments. Section~\ref{sec:discussion} is a further discussion of implications of this work on the stability of the AMOC for rate-dependent perturbed CO${}_2$ scenarios. 

\subsection{Box models of the AMOC}
\label{sec:concept}

The two-box model of Stommel \cite{Stommel1961} can be written in non-dimensional form as
\begin{eqnarray}
\frac{dT}{dt}&=\eta_1-T(1+|T-S|) \label{eq:twobox},\\
\frac{dS}{dt}&=\eta_2-S(\eta_3+|T-S|)\nonumber,
\end{eqnarray}
and it is an idealisation of the dynamical interaction between non-dimensional equator-to-pole temperature gradient $T$ and a corresponding salinity gradient $S$ \cite{Dijkstra2013}. There are three non-negative dimensionless parameters: $\eta_1$ and $\eta_2$ represent the relative strength of thermal and freshwater forcing and $\eta_3$ is the ratio of thermal to freshwater surface restoring times. Note that even though this system is composed of two coupled ODEs where each expression is quadratic, there remain fundamental unsolved problems about the dynamics of quadratic systems on the plane - in particular Hilbert's 16th problem is unsolved - this conjectures that there can only be finitely many limit cycles for such a system \cite{Dumortier2000}.

\begin{figure}
\centering
\includegraphics[width=0.6\textwidth]{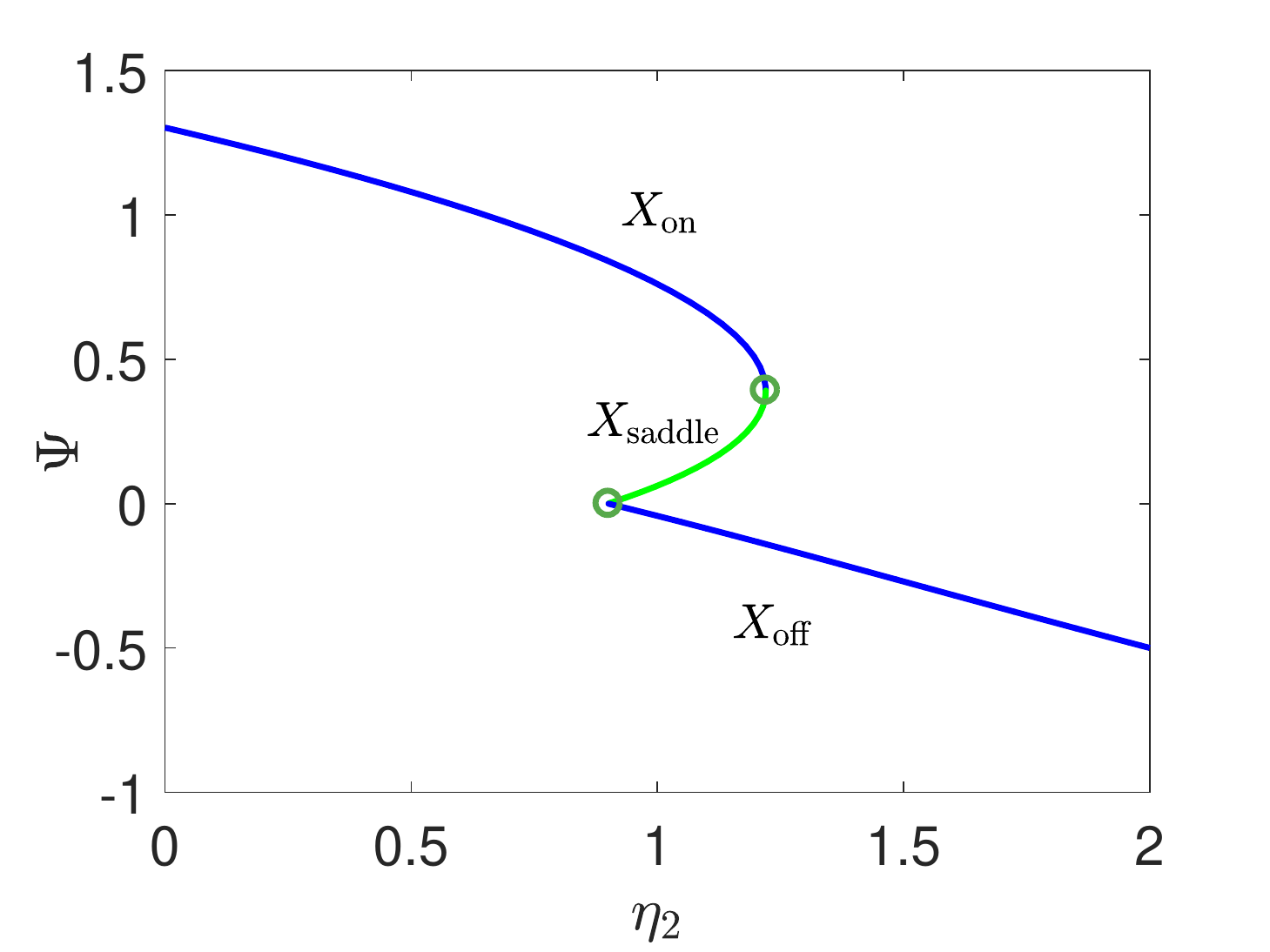}

\caption{ Bifurcation diagram showing $\Psi=T-S$ plotted against $\eta_2$ for $\eta_1=3.0$ and $\eta_3=0.3$ for the two-box model (\ref{eq:twobox}): see for example \cite[Figure 10.4]{Dijkstra2013}. Observe the upper and lower branches correspond to AMOC ``on'' and ``off'' states. Note the saddle node bifurcations at the end of each branch, the lower of which is a non-smooth saddle node. 
\label{fig:twoboxbifs}
}
\end{figure}

As recognised already by Stommel \cite{Stommel1961}, the system (\ref{eq:twobox}) allows two stable equilibrium states, and bifurcations (possibly non-smooth, due to the terms $|T-S|$ in (\ref{eq:twobox}) that are continuous but not differentiable at $T=S$ \cite{Bernardo2008,Kowalczyk2011}) that go between these states on varying a parameter (typically $\eta_2$). 
On varying $\eta_2$ in (\ref{eq:twobox}) we can find a bifurcation diagram as shown in Figure~\ref{fig:twoboxbifs} which shows a region of hysteresis with two stable states. Note that the linear stability of any equilibrium $(T^*,S^*)$ for (\ref{eq:twobox}) with $T^*\neq S^*$ and $\phi= \mathrm{sgn}(T^*-S^*)$ is determined by the Jacobian
\begin{equation}
J=\begin{bmatrix}
-1+[-2T^*+S^*]\phi & T^*\phi\\
-S^*\phi & -\eta_3+[2S^*-T^*]\phi
\end{bmatrix}.
\end{equation}
Note that
$$
\mathrm{Tr}(J)=-1-\eta_3+[-T^*+S^*]\phi= -1-\eta_3-3|T^*-S^*|.
$$
The fact that this trace is negative for all positive $\eta_3$ implies that there can be no Hopf bifurcation of an equilibria for (\ref{eq:twobox}): the only smooth local bifurcations that occur will have one dimensional centre manifold, in this case saddle-node. There are also non-smooth saddle-node bifurcations of equilibria with $T^*=S^*$ \cite{Kowalczyk2011}. Figure~\ref{fig:twoboxbifs} shows a typical bifurcation diagram on varying $\eta_2$ for this model: note that complex eigenvalues can appear at equilibria even though there is no Hopf bifurcation.

Modifications of this two-box model can be reduced to an It\^{o} SDE for one variable $y(t)$ \cite[Equations 10.19, 10.20]{Dijkstra2013} of the form
\begin{equation}
dy=-V'(y,F)dt+\sigma dW_t,
\label{eq:twobox_reduced}
\end{equation}
where 
$$
V(y,F)=-Fy+y^2/2+\mu^2(y^4/4-2y^3/3+y^2/2)
$$
is a potential with two wells corresponding to AMOC ``on" and ``off" states, $F$ represents an (possibly  time-dependent) injection of a freshwater flux in the North Atlantic box corresponding to a ``hosing" of this with fresh water, and $W_t$ is a noise process with amplitude $\sigma$ that represents a stochastic component to the freshwater flux \cite{cessi1994simple}.  The model (\ref{eq:twobox_reduced}) for $\sigma=0$ effectively produces the same behaviour as (\ref{eq:twobox}) and a bifurcation diagram similar to Figure~\ref{fig:twoboxbifs}, namely a region of hysteresis between the ``on'' and ``off'' states, $X_{\mathrm{on}}$ and $X_{\mathrm{off}}$, terminated by saddle-node bifurcations with the unstable separating state $X_{\mathrm{saddle}}$. More complex models, such as the multiple box models studied in \cite{LucariniStone2005,Titzetal2002b,Titzetal2002a,LucariniFaranda2012} 
show richer dynamics including Hopf bifurcations.

\section{Bifurcations of an AMOC box model}
\label{sec:fivebox}

\begin{figure}
\centering
\includegraphics[width=0.6\textwidth]{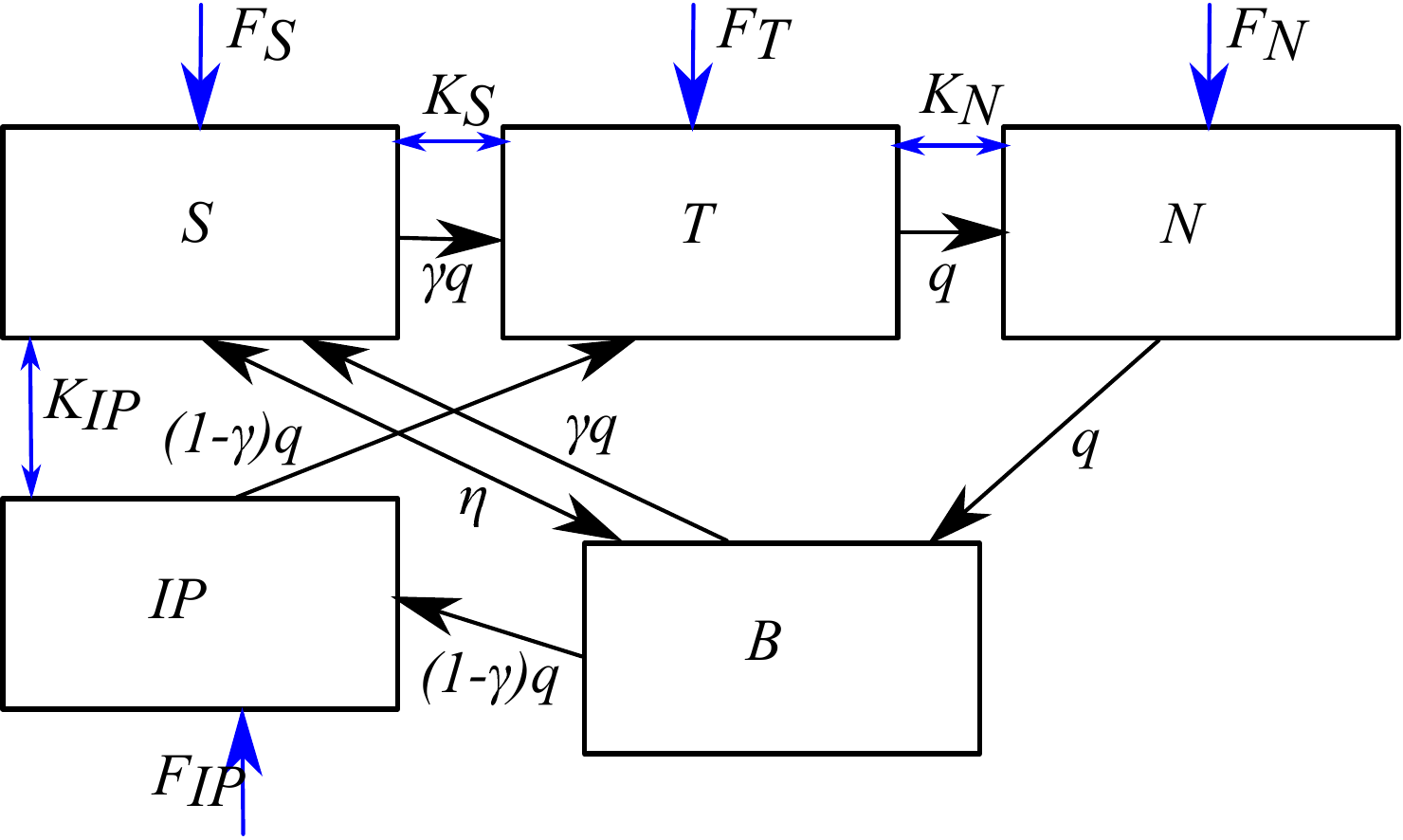}
\caption{
The flows between the five boxes representing (from top left) Southern Ocean, Tropical Atlantic, North Atlantic, Indo-Pacific and bottom waters, in the global ocean box model of \cite{RWSH17} (cf their Fig.~1). There are freshwater surface fluxes $F_X$ and exchange of salinity via the indicated arrows. The $K_X$ indicate wind-driven fluxes while the advective flux due to the AMOC, $q$, is driven by a balance of salinity and temperature gradients. There is an $S$/$B$ mixing parameter $\eta$. A proportion $\gamma$ of the flux $q$ follows a cold water path via the Southern Ocean, while the remainder follows the warm water path via the Indo-Pacific.}\label{fig:fiveboxflows}
\end{figure}

By examining the geometry and behaviour of ocean currents and salinity for the experiments of Hawkins et al \cite{hawkins2011} using the FAMOUS AOGCM \cite{Smith2012}, Wood et al \cite{RWSH17} propose a realistic box model with five boxes that models the balance of salinity in the main ocean water masses. They consider North Atlantic ($N$) and Tropical Atlantic ($T$) boxes, as well as an Indo-Pacific ($IP$) box, and assume that the temperature of the North Atlantic box is dependent on the AMOC strength. These boxes are coupled together via boxes corresponding to the Southern Ocean ($S$) and the bottom waters ($B$). Figure~\ref{fig:fiveboxflows} schematically illustrates the salinity fluxes between boxes.

The five box model is as follows. We assume that the AMOC flow $q$ is proportional to the density gradient between the $N$ and $S$ boxes (which in turn is determined by the temperatures and salinities of the boxes):
$$
q=\frac{\lambda [\alpha (T_S - T_0) + \beta (S_N - S_S)]}{1+\lambda\alpha\mu}.
$$
For $q \geq 0$ conservation of salt for the five boxes gives:

\begin{equation}
\label{eq:AOMC_model_q>0}
\left.\begin{array}{rcl}
V_N \frac{dS_N}{dt}	 &=& q(S_T - S_N) + K_N (S_T - S_N) - F_N S_0, \\
V_T \frac{dS_T}{dt}	&=& q [\gamma S_S + (1- \gamma) S_{IP} - S_T] + K_S (S_S - S_T)+K_N(S_N - S_T) - F_T S_0,\\
V_S \frac{dS_S}{dt}	&=& \gamma q (S_B -S_S) + K_{IP} (S_{IP} - S_S) + K_S (S_T - S_S) +\eta (S_B - S_S) - F_S S_0, \\
V_{IP} \frac{dS_{IP}}{dt} &=& (1-\gamma)q(S_B - S_{IP}) + K_{IP} (S_S - S_{IP}) - F_{IP} S_0, \\
V_{B} \frac{d S_B}{dt} &=& q(S_N - S_B) + \eta (S_S - S_B),
\end{array}\right\}
\end{equation}
while for $q < 0$ we have
\begin{equation}
\label{eq:AOMC_model_q<0}
\left.\begin{array}{rcl}
V_N \frac{dS_N}{dt}	 &=& |q|(S_B - S_N) + K_N (S_T - S_N) - F_N S_0, \\
V_T \frac{dS_T}{dt}	&=& |q| (S_N - S_T) + K_S (S_S - S_T)+K_N(S_N - S_T) - F_T S_0,\\
V_S \frac{dS_S}{dt}	&=& \gamma |q| (S_T -S_S) + K_{IP} (S_{IP} - S_S) + K_S (S_T - S_S) +\eta (S_B - S_S) - F_S S_0, \\
V_{IP} \frac{dS_{IP}}{dt} &=& (1-\gamma)|q|(S_T - S_{IP}) + K_{IP} (S_S - S_{IP}) - F_{IP} S_0, \\
V_{B} \frac{d S_B}{dt} &=& \gamma |q| S_S + (1-\gamma) |q| S_{IP} - |q| S_B + \eta (S_S -S_B).
\end{array}\right\}
\end{equation}

If we consider the total salt content $C$,
\begin{equation}
C= V_N S_N+ V_T S_T+ V_S S_S+ V_{IP} S_{IP} + V_B S_B,
\label{eq:totalsalt}
\end{equation}
then 
$$
dC/dt=-(F_N+F_T+F_S+F_{IP})S_0.
$$
So if the total surface fresh water fluxes satisfy
\begin{equation}
F_N+F_T+F_S+F_{IP}=0,
\label{eq:fluxesbalance}
\end{equation}
then $C$ is constant on trajectories of (\ref{eq:AOMC_model_q>0},\ref{eq:AOMC_model_q<0}). This means that we can eliminate one of the equations. 

Note that if we assume $C$ is constant and solve (\ref{eq:totalsalt}) to give $S_B$ in terms of the other variables this corresponds to the solutions of the original system as long as (\ref{eq:fluxesbalance}) is satisfied. If we impose (\ref{eq:totalsalt}) but (\ref{eq:fluxesbalance}) is not satisfied, then we are effectively solving the system with an additional flux $F_B$ to the $V_B\frac{dS_B}{dt}$ equation, such that $F_B=-F_N-F_S-F_T-F_{IP}$. 

In this paper we consider values of the model parameters that are calibrated to the $FAMOUS_B$ AOGCM model version \cite{Smith2012} with either pre-industrial $\mathrm{CO}_2$ concentrations (``$1 \times \mathrm{CO}_2$'', shown in Appendix~\ref{sec:appdx1}, Table~\ref{tab:oceans} and Table~\ref{tab:params}), or with doubled $\mathrm{CO}_2$ concentrations (``$2 \times \mathrm{CO}_2$'', shown in Appendix~\ref{sec:appdx1}, Table~\ref{tab:oceans2CO2} and Table~\ref{tab:params2CO2}). See \cite{RWSH17} for details.

\subsection{A three box reduction of the five box model}

We consider an empirical three box reduction as follows. We note from \cite[Figure 2]{RWSH17} that the variation of $S_S$ and $S_B$ are apparently small and slow compared to $S_N$ and $S_T$. Hence we fix $S_S$ and $S_B$ and consider only $S_N$, $S_T$ and $S_{IP}$ as dynamic variables. Conservation of salt (\ref{eq:totalsalt}) means we can solve for one of the variables (we choose $S_{IP}$) to give
\begin{equation}
\label{eq:AOMC_model_q>0_3box}
\left.\begin{array}{rcl}
V_N \frac{dS_N}{dt}	 &=& q(S_T - S_N) + K_N (S_T - S_N) - F_N S_0, \\
V_T \frac{dS_T}{dt}	&=& q [\gamma S_S + (1- \gamma) S_{IP} - S_T] + K_S (S_S - S_T)+K_N(S_N - S_T) - F_T S_0,\\
\end{array}\right\}
\end{equation}
for $q\geq 0$ and
\begin{equation}
\label{eq:AOMC_model_q<0_3box}
\left.\begin{array}{rcl}
V_N \frac{dS_N}{dt}	 &=& |q|(S_B - S_N) + K_N (S_T - S_N) - F_N S_0, \\
V_T \frac{dS_T}{dt}	&=& |q| (S_N - S_T) + K_S (S_S - S_T)+K_N(S_N - S_T) - F_T S_0,\\
\end{array}\right\}
\end{equation}
for $q<0$, where we fix $S_S$, $S_B$ and use (\ref{eq:totalsalt}) to determine the remaining variable $S_{IP}$. The baseline parameters (we refer to as $1\times \mathrm{CO}_2$ as they are tuned to FAMOUS runs with pre-industrial atmospheric $\mathrm{CO}_2$) are given in Appendix~\ref{sec:appdx1}.  For the numerical studies presented in the following sections we will use scaled salinities to enable more accurate numerical computation: $\tilde{S}_i = 100(S_i - S_0)$ for $i\in\{N,T,S,IP,B\}$ (see Appendix~\ref{sec:appdx2} for the modified equations). To convert to the commonly used units of parts per thousand, $S_{ppt} = 35.0 + 10\tilde{S}$.

\subsection{Bifurcations of the five box model}

We consider variation of the fresh water fluxes as in \cite{RWSH17}: we consider a ``hosing'' forcing $H$ which may depend on time and that affects all of the fresh water fluxes simultaneously.
 The pattern of hosing represents a surface fresh water input into the North Atlantic over 20-50$^{o}$N (into parts of both $N$ and $T$ boxes) and removal elsewhere in order to preserve fresh water content so that (\ref{eq:fluxesbalance}) is satisfied, as in \cite{hawkins2011}. Table~\ref{tab:hosingfluxes} shows the surface fluxes corresponding to hosing $H$, for the $1\times \mathrm{CO}_2$ parameter set. 

\begin{table}
$$
\begin{array}{l|rcl|r}
F_N & 0.384+ 0.1311 H &~~~& F_S &  1.078-0.2626 H\\
F_T & -0.723+0.6961 H &~~~& F_{IP} & -0.738-0.5646 H
\end{array}
$$
\caption{Surface freshwater fluxes (all in $\Sv$) as a function of the ``hosing" variable $H$, with the relationship shown here for the $1\times\mathrm{CO}_2$ parameter set. The total flux is zero for all $H$ and corresponds to the baseline values in Table~\ref{tab:oceans} for $H=0$. }
\label{tab:hosingfluxes}
\end{table}

\begin{figure}
\centering

	{\includegraphics[width = \textwidth]{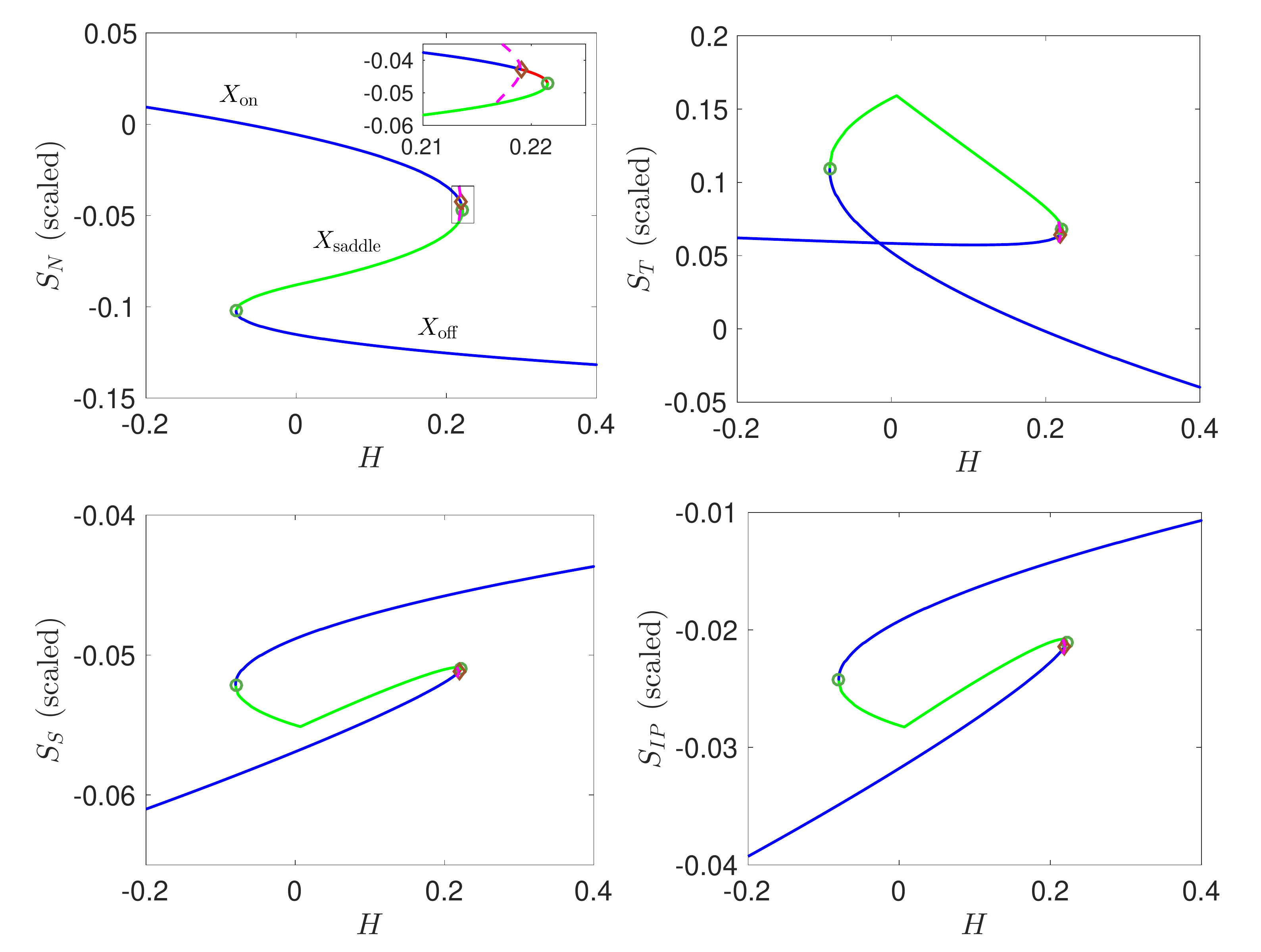}}
	
	\caption[The bifurcation diagram]{The bifurcation diagram of the five box model (\ref{eq:AOMC_model_q>0},\ref{eq:AOMC_model_q<0}) on varying $H$ ($1\times \mathrm{CO}_2$ parameters). The stable equilibria are shown in blue, unstable in green. The green circles indicate saddle node bifurcations, and the brown diamonds indicate Hopf bifurcations.  The dotted pink line shows the minimum and maximum of the unstable periodic orbit.  A zoom of the upper bifurcation region is shown for $S_N$. Note that 0.1 scaled salinity units as shown here corresponds to a salinity difference of $1 ppt$ in oceanographic units, with a scaled salinity of $0$ corresponding to an absolute value of $35 ppt$. }
	    \label{fig:fiveboxbifs1CO2}
 
\end{figure}

\begin{figure}
\centering
 {\includegraphics[width=0.8\textwidth]{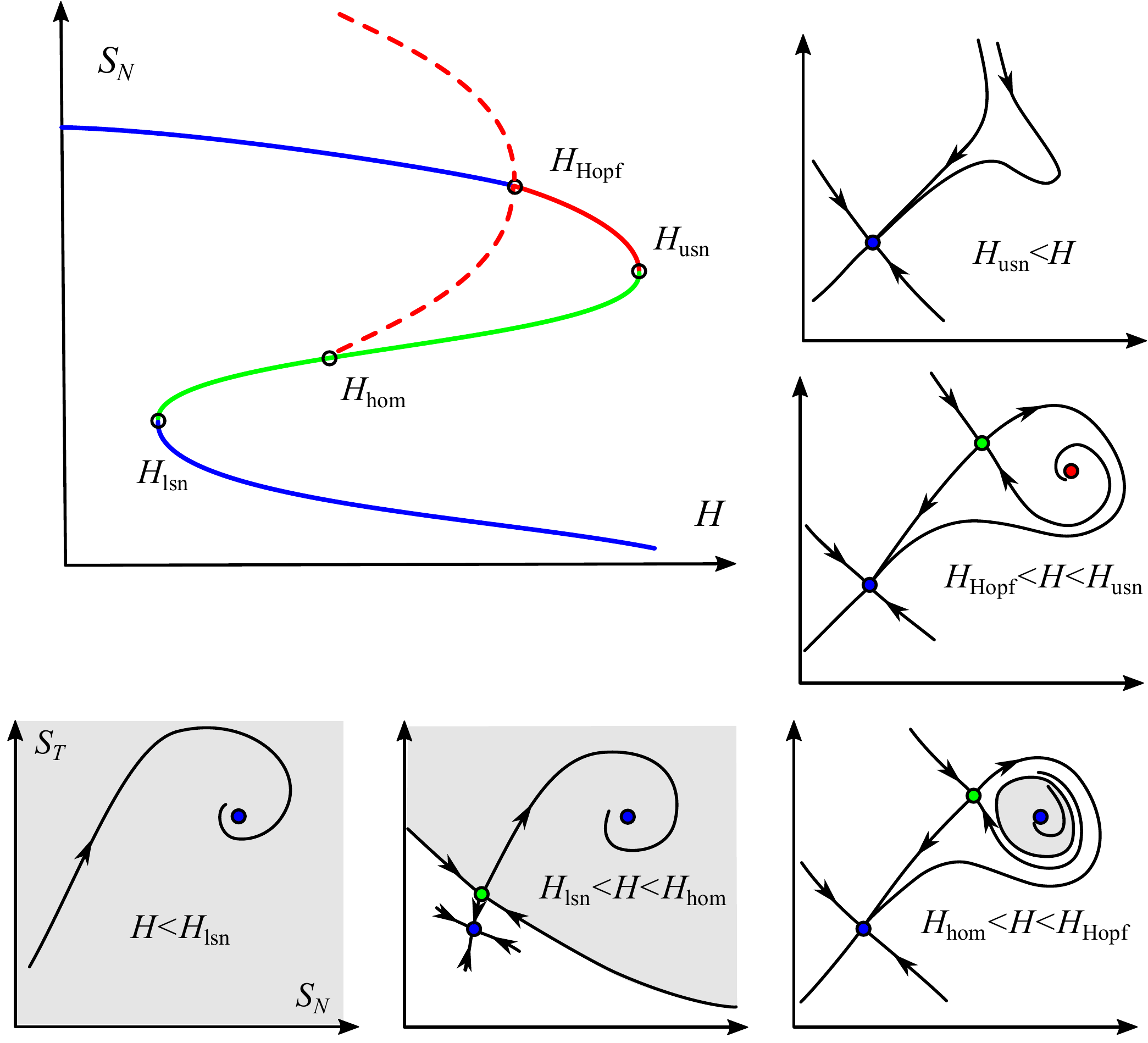}}
 
\caption{Schematic of the bifurcation diagram (top left) and phase portraits on varying $H$ through various points on the bifurcation diagram of five box and three box models (\ref{eq:AOMC_model_q>0_3box},\ref{eq:AOMC_model_q<0_3box}) for both $1\times \mathrm{CO}_2$ and $2\times \mathrm{CO}_2$ parameters. The shaded region shows the basin of attraction for the upper branch - this shrinks from being semi-infinite size for $H<H_{\mathrm{hom}}$ to being finite size for $H>H_{\mathrm{hom}}$.
As in Figure~\ref{fig:fiveboxbifs1CO2}, blue branches are stable, red are unstable  and green are saddles. Note that the upper branch loses stability before the saddle node.  The dotted lines represent minimum and maximum values of periodic orbits. 
\label{fig:threebox_sketchbifs} }
\end{figure}

We present xppaut \cite{xppautref} and COCO \cite{cocoref} continuation and bifurcation analysis of (\ref{eq:AOMC_model_q<0},\ref{eq:AOMC_model_q>0}) for the $1\times\mathrm{CO}_2$ scenario. Similarly to the two-box model, there is a hysteresis between two branches of stable equilibria $X_{\mathrm{on}}$ and $X_{\mathrm{off}}$, separated by saddles $X_{\mathrm{saddle}}$, on varying $H$ and imposing $C$ constant (ie. using (\ref{eq:totalsalt}) to specify $S_{IP}$), as shown in Figure~\ref{fig:fiveboxbifs1CO2}. Figure~\ref{fig:threebox_sketchbifs} schematically shows changes to the system on varying $H$. The behaviour is qualitatively the same for either $1\times\mathrm{CO}_2$ or $2\times\mathrm{CO}_2$ parameters. The hysteresis in this model is not as straightforward as appears in the two-box model but is more like that found in \cite{Titzetal2002a,Titzetal2002b}.

In particular, although the lower branch is destroyed at a saddle- node bifurcation, the upper branch loses stability at a subcritical Hopf bifurcation before it is destroyed. In addition there is a homoclinic bifurcation (see Figure \ref{fig:fiveboxbifs1CO2}) that leads to a sudden collapse in the size of the basin of attraction before the Hopf bifurcation occurs.

\subsection{Bifurcations of the three box model}

The bifurcation structures for the five box model are qualitatively the same as those of the three box reduction (\ref{eq:AOMC_model_q>0_3box},\ref{eq:AOMC_model_q<0_3box}) on varying $H$, and both are qualitatively the same for both $1\times \mathrm{CO}_2$ and $2\times \mathrm{CO}_2$ parameters. Indeed, Table~\ref{tab:comparebifsH} compares the locations of the four bifurcations for the two systems. With the assistance of computer algebra (Maple) it is possible to find all branches for $q\geq 0$ and $q<0$ respectively to arbitrary precision and the local (i.e. saddle-node and Hopf) bifurcation points for the three box model: these parameter values are included on Table~\ref{tab:comparebifsH}. The homoclinic bifurcations are approximated by following a large but fixed period orbit near the end of the branch that bifurcates from the Hopf bifurcation.

\begin{figure}
\centering
 {\includegraphics[width=0.5\textwidth]{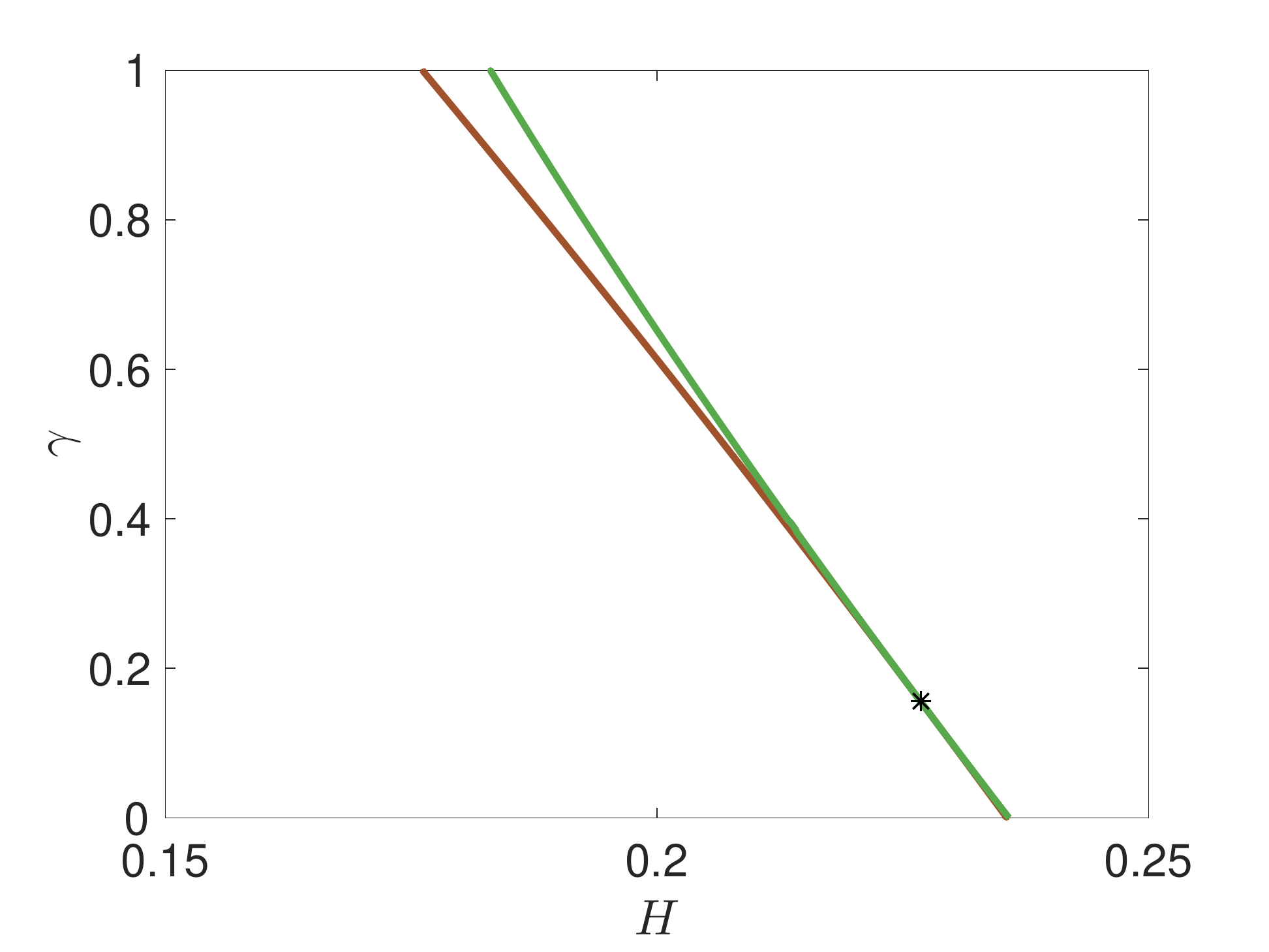}}
 
\caption{Two parameter plot showing the location of $H_{\mathrm{Hopf}}$ (red) and $H_{\mathrm{usn}}$ (green) for the three box model ($1\times \mathrm{CO}_2$ parameters). Other details are as in Figure~\ref{fig:threeboxbifs2CO2} on varying $\gamma$ away from the default value $0.39$. The two curves meet at a Bogdanov-Takens point at $(H,\gamma)= (0.2268,0.1559)$ while the Hopf and upper saddle node bifurcations move apart for larger $\gamma$ and smaller $H$.}     
\label{fig:threebox_2parambif}
\end{figure}

Comparing the bifurcations of the five and three box models on varying $H$, note there is close agreement qualitatively (and quantitatively for $1\times \mathrm{CO}_2$): see Table~\ref{tab:comparebifsH}. The bifurcation scenario is robust: for example on changing the parameter $\gamma$ we find the Hopf curve meets the upper saddle node bifurcations on reducing $\gamma$: see Figure~\ref{fig:threebox_2parambif}. The two curves meet at a Bogdanov-Takens (BT) point at $(H,\gamma)=(0.2268,0.1559)$ that can be determined (with the aid of Maple) by simultaneous solution of an equilibrium of (\ref{eq:AOMC_model_q>0_3box}) with $q>0$ whose Jacobian has both determinant and trace zero. Note that the BT point represents a convergence not just of Hopf and Saddle node bifurcations but also we can infer \cite{Kuznetsov1998} that a homoclinic bifurcation continues to the BT point.


\begin{table}
$$
\begin{array}{r|rrr|rrr}
& & \mbox{$1\times \mathrm{CO}_2$} & & & \mbox{$2\times \mathrm{CO}_2$}& \\
& \mbox{Five box} & \mbox{Three box}  & \mbox{Three box} & \mbox{Five box} &\mbox{Three box}&\mbox{Three box}\\
 & \mbox{numerical} & \mbox{numerical} & \mbox{Maple} & \mbox{numerical} & \mbox{numerical}&\mbox{Maple}\\
\hline
H_{\mathrm{lsn}} & -0.07996  & -0.05446 & -0.05445 & -0.4056 & -0.3795& -0.3792 \\
H_{\mathrm{hom}} & 0.2165 & 0.2128 & \mbox{NA}& 0.4510 & 0.3555 & \text{NA} \\
H_{\mathrm{Hopf}} & 0.2191   & 0.2134 &  0.2133& 0.4789 & 0.3895& 0.3888 \\
H_{\mathrm{usn}}  & 0.2214  & 0.2139 & 0.2138& 0.511 & 0.4236 &  0.4225 \\
\end{array}
$$
\caption{Comparison of the locations of bifurcations for $H$ for the five box (\ref{eq:AOMC_model_q>0},\ref{eq:AOMC_model_q<0}) and the three box (\ref{eq:AOMC_model_q>0_3box},\ref{eq:AOMC_model_q<0_3box}) model. Note that $H_{\mathrm{lsn/usb}}$ are the saddle-node (fold) bifurcations on the lower and upper branches respectively, $H_{\mathrm{hom/Hopf}}$ are respectively the homoclinic and Hopf bifurcations on the upper branch: see Figure~\ref{fig:threebox_sketchbifs}. The numerical solutions by path-following use COCO, apart from the homoclinic bifurcations which found using xppaut. The last column gives the analytically found values for the three box model using Maple (not available for $H_{hom}$). 
}

\label{tab:comparebifsH}
\end{table} 

\begin{figure}
\centering
	{\includegraphics[width = \textwidth]{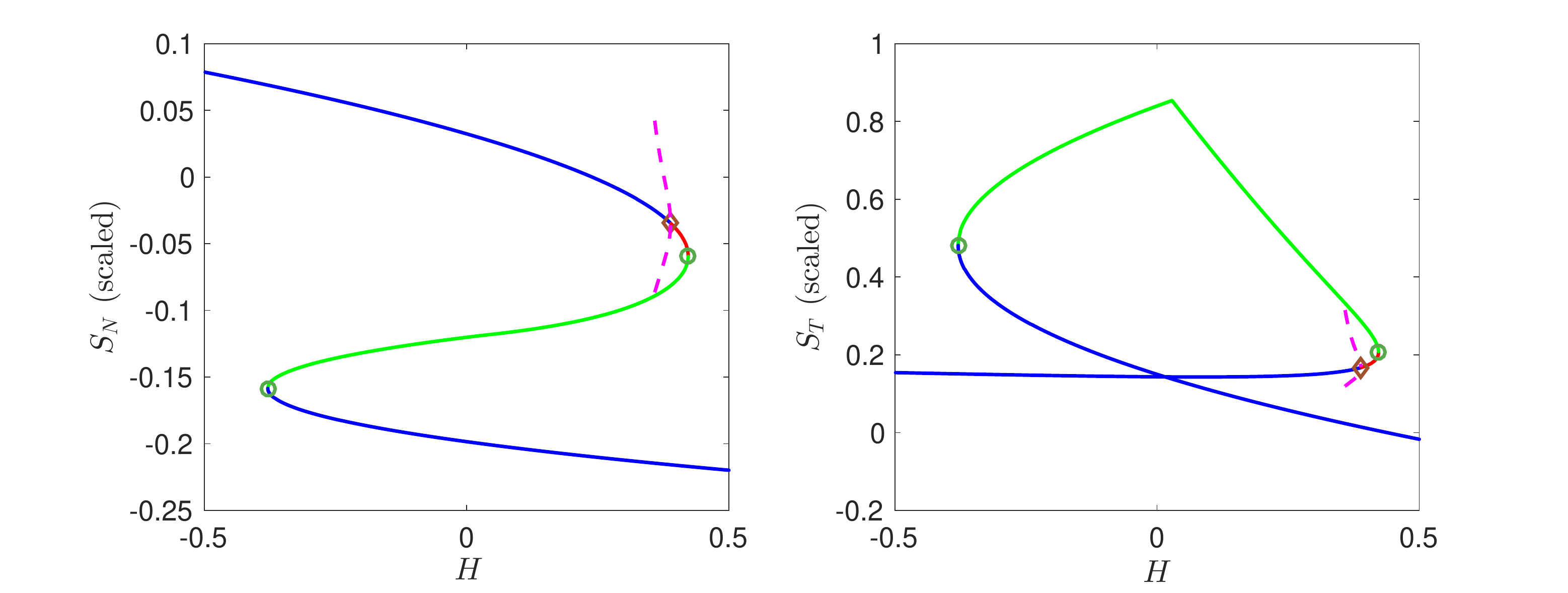}}
	\caption[The bifurcation diagram]{The bifurcation diagram of the three box model (\ref{eq:AOMC_model_q>0_3box},\ref{eq:AOMC_model_q<0_3box}) when parameters are tuned to atmospheric conditions of doubled preindustrial CO$_2$ ($2\times \mathrm{CO}_2$). Observe an equivalent set of bifurcations but a larger area of instability before the saddle-node as compared to the $1\times \mathrm{CO}_2$ case shown in Figure~\ref{fig:fiveboxbifs1CO2}.}
	    \label{fig:threeboxbifs2CO2}

\end{figure}

 Table~\ref{tab:comparebifsH} and Figure~\ref{fig:threeboxbifs2CO2} also show the bifurcations for the 3-box model with the $2\times \mathrm{CO}_2$ parameter set. We see that for $2\times \mathrm{CO}_2$ there is a larger separation (in $H$) between the Hopf and upper saddle node bifurcations. For this reason we use the $2\times \mathrm{CO}_2$ parameters from here on, for clarity of exposition. We expect qualitatively similar results for the $1\times \mathrm{CO}_2$ parameters.

\subsection{Basin bifurcations}

\begin{figure}
\centering
    	\subcaptionbox{} [0.45\textwidth]
    {\includegraphics[width=0.45\textwidth]{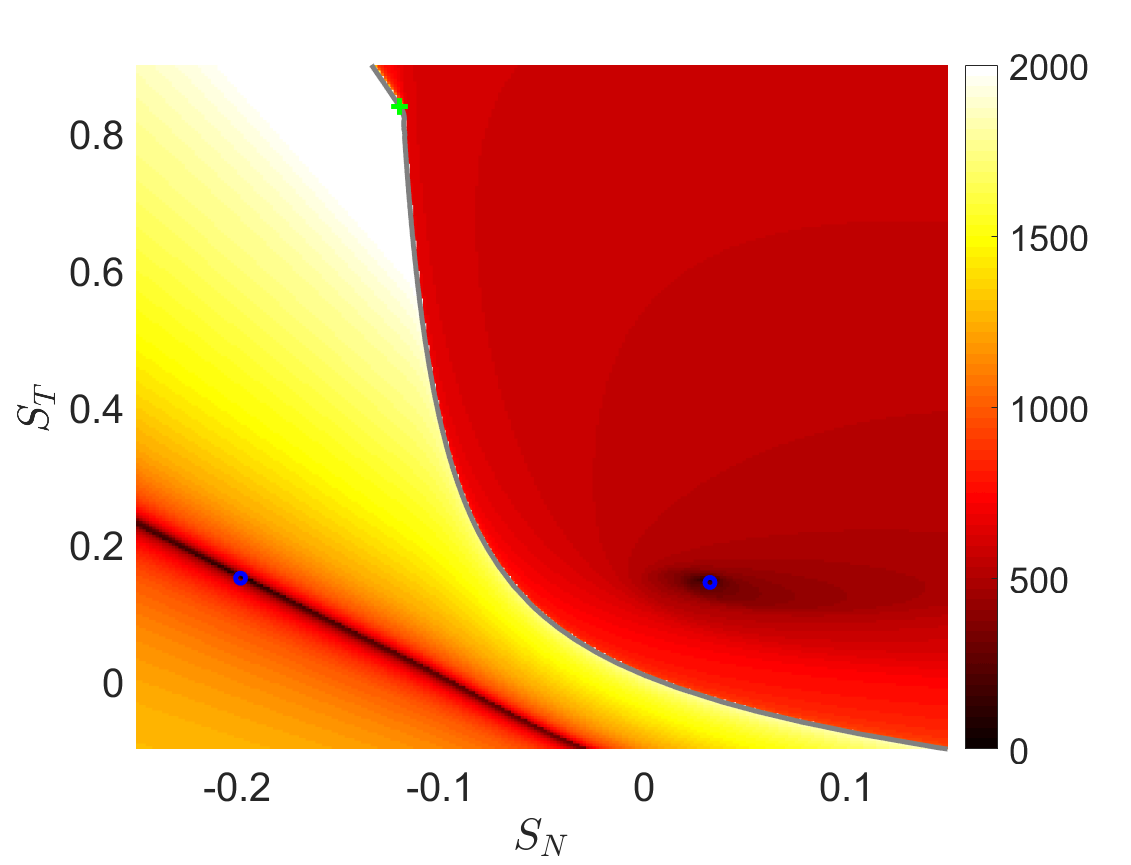}}
    	\subcaptionbox{} [0.45\textwidth]
    {\includegraphics[width=0.45\textwidth]{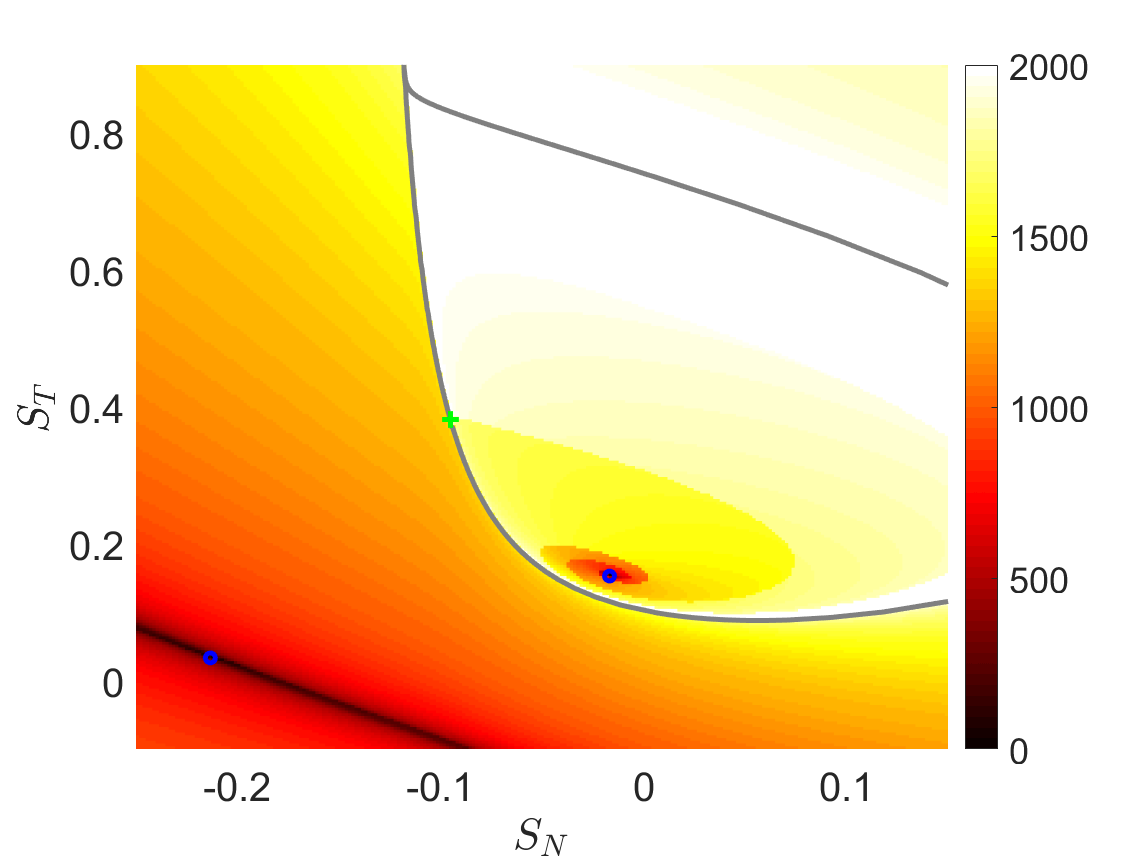}}
    	\subcaptionbox{} [0.45\textwidth]
    {\includegraphics[width=0.45\textwidth]{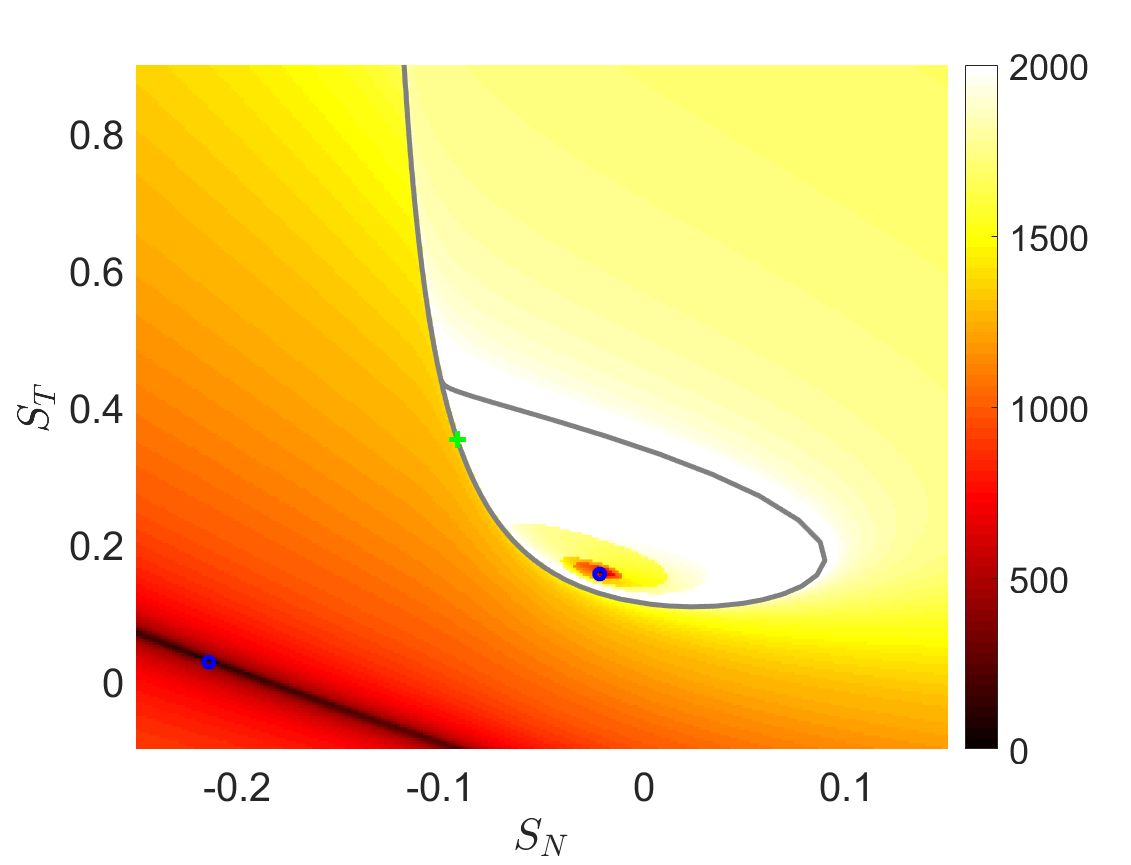}}
    	\subcaptionbox{} [0.45\textwidth]
    {\includegraphics[width=0.45\textwidth]{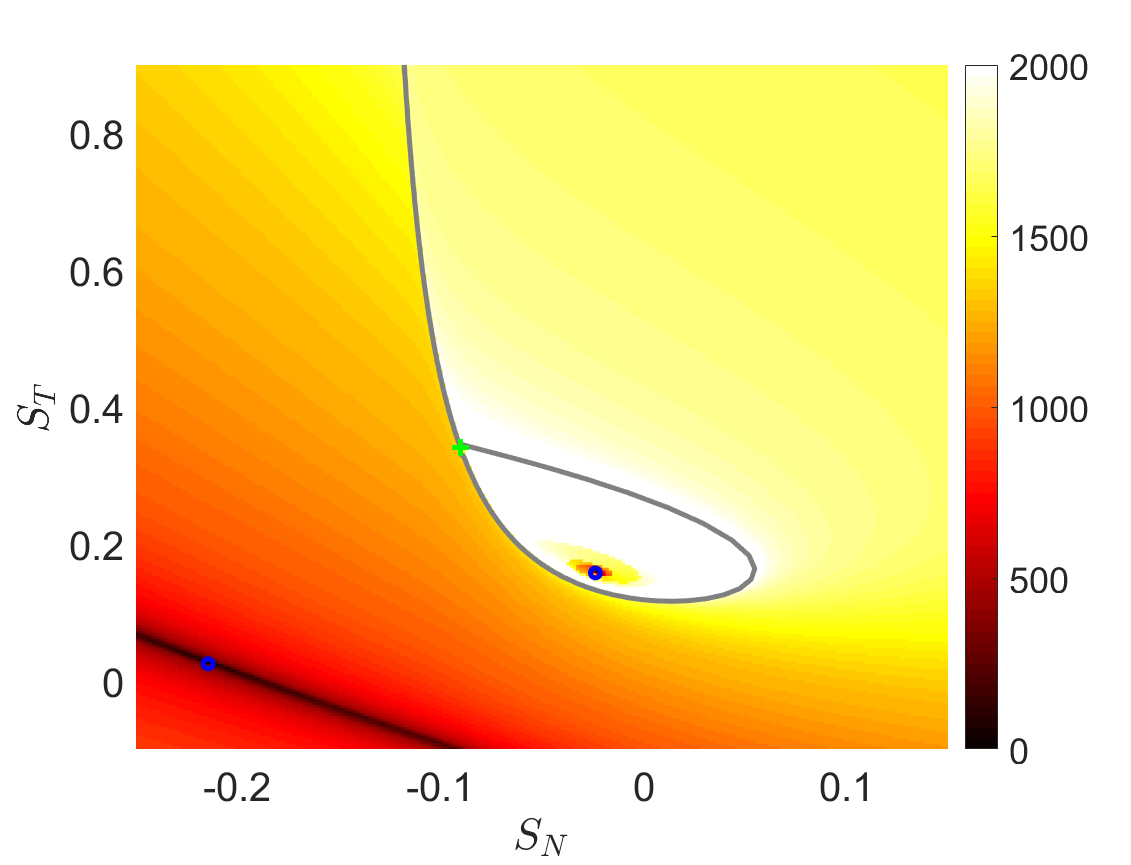}}
 \caption{ The basins of attraction of the two stable equilibria for different values of $H$, system (\ref{eq:AOMC_model_q>0_3box},\ref{eq:AOMC_model_q<0_3box}) ($2\times \mathrm{CO}_2$ parameters). The colour map represents the duration (in years) to get as close as $10^{-3}$ to the equilibria. The values of $H$ are (a) 0, (b) 0.32678, (c) 0.34696 and (d) 0.35503. The integration for basins of attraction use 4th order Runge-Kutta with stepsize $10$ years. {\em Supplementary Movie {\tt MOC\_basin\_2co2.mp4} shows an animation of this figure.}}
 
 \label{fig:threebox_basins_H}
\end{figure}

Figures~\ref{fig:threebox_sketchbifs} and \ref{fig:threebox_basins_H} shows changes that take place in the basin of attraction of the upper branch $X_{\mathrm{on}}$ on varying $H$ (3-box model, $2\times \mathrm{CO}_2$ parameters). We observe that (similarly to \cite{Titzetal2002a,Titzetal2002b})
\begin{itemize}
\item The approach to the upper branch is oscillatory: for the entire stable part of the upper branch there is a complex pair of slowly attracting eigenvalues for the linearisation. These become real at a spiral to node transition between Hopf and Saddle-node points.
\item The basin of the upper branch shrinks from unbounded to a small and finite size at a homoclinic bifurcation just before the Hopf bifurcation.
\end{itemize}
This has some counter-intuitive effects for perturbations, for example it predicts an oscillatory overshoot of $S_N$ at a perturbation that switches the system from the lower to upper branch. As the basin of the upper branch shrinks to a small size, small perturbations in any direction can take the system out of the basin of the upper branch.

\section{Tipping in response to time-dependent forcing}
\label{sec:tipping}

In this section we illustrate the potential tipping behaviour of the AMOC in response to time-dependent forcing. We use the 3-box model with the $2\times \mathrm{CO}_2$ parameter set and integrate using a 4th order Runge-Kutta scheme with stepsize between $1$ and $2$ years.

\subsection{Time-dependent hosing}

We consider the influence of time-dependent perturbations of (\ref{eq:AOMC_model_q>0_3box},\ref{eq:AOMC_model_q<0_3box}) where the freshwater fluxes $F_X$ are varied via $H(t)$ according to a given protocol of variation that allows exploration of various scenarios of moving around the bifurcation diagram. The time-dependent hosing can be considered as an idealised representation of different future climate change mitigation scenarios, or palaeo-climatic periods.

If $H(t)$ is varied slowly enough, standard arguments of adiabatic reduction \cite{Kuehn2015} can be used to predict the behaviour of the nonautonomous perturbed system. However, it is not always clear how slow this needs to be, and in many cases there are scientifically important questions that need exploration of a range of fast or slow perturbations; for example current anthropogenic forcing of the climate system is on a similar or faster timescale than the inherent timescales of AMOC adjustment. 

The particular protocol we consider here is inspired by the question of how quickly changes to hosing need to be reversed to avoid tipping of the upper branch onto the lower branch \cite{Ritchie2019}. This represents an idealisation of a scenario in which climate forcing temporarily overshoots its final value.

\begin{figure}
\centering
\includegraphics[width=0.7\textwidth]{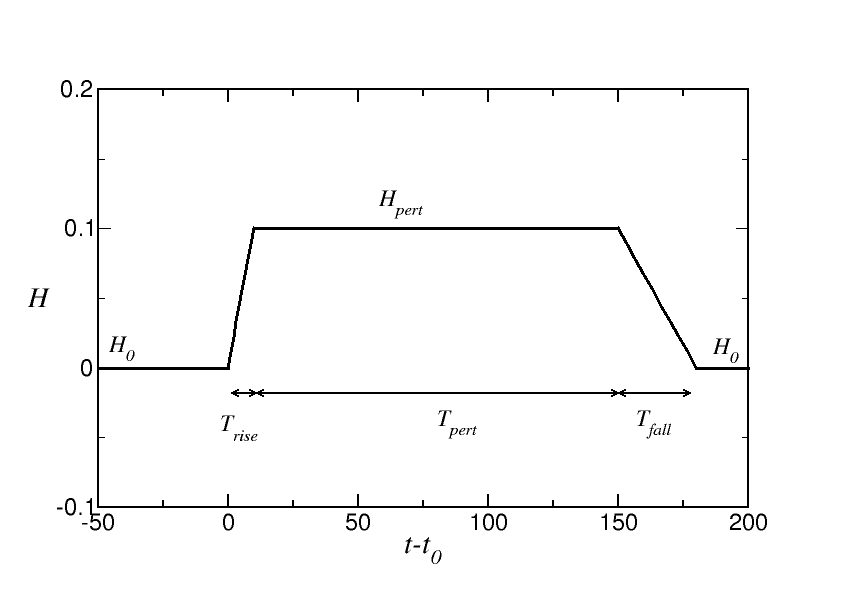}
\caption{Schematic diagram showing a piecewise linear time-dependent perturbation $H(t)=H_{\mathrm{pwl}}(t-t_0)$ where $t_0=50$, $T_{\mathrm{rise}}=10$, $T_{\mathrm{pert}}=140$, $T_{\mathrm{fall}}=30$, $H_0=0$ and $H_{\mathrm{pert}}=0.1$. Whether the system tips from one state to the other depends not only on $H_{\mathrm{pert}}$ but also the other parameters. In the special case $T_{\mathrm{rise}}=T_{\mathrm{fall}}=0$ we refer to the perturbation as a ``press''.}
\label{fig:PWL}
\end{figure}

To this end, we consider the effect of piecewise linear (PWL) perturbations $H(t)=H_{pwl}(t-t_0)$ (visualized in Figure~\ref{fig:PWL}, ) on the system, where
\begin{equation}
H_{pwl}(t) =\begin{cases} H_0 & t<0, \\
		\alpha(t) & t\in[0,T_{\mathrm{rise}}], \\
        H_{\mathrm{pert}} & t-T_{\mathrm{rise}}\in[0,T_{\mathrm{pert}}],\\
        \beta(t) & t-T_{\mathrm{rise}}-T_{\mathrm{pert}}\in[0,T_{\mathrm{fall}}], \\
        H_0 & t\geq T_{\mathrm{rise}}+T_{\mathrm{pert}}+T_{\mathrm{fall}},
        \end{cases}
\label{eq:H_PWL}
\end{equation}
where $\alpha(t)$ and $\beta(t)$ are linear functions such that $H$ is continuous. This depends on constants $T_{\mathrm{rise}}$, $T_{\mathrm{pert}}$, $T_{\mathrm{fall}}$, and levels $H_0$ and $H_{\mathrm{pert}}$. If we define rise and fall rates
\begin{equation}
r_{\mathrm{rise}}=\frac{|H_0-H_{\mathrm{pert}}|}{T_{\mathrm{rise}}},~
r_{\mathrm{fall}}=\frac{|H_0-H_{\mathrm{pert}}|}{T_{\mathrm{fall}}},
\end{equation}  
then note that $\alpha(t)=r_{\mathrm{rise}}t$ and $\beta(t)=r_{\mathrm{fall}}(t-T_{\mathrm{rise}}-T_{\mathrm{pert}})$.  Similar PWL forcing experiments to the AMOC are explored in \cite{ LucariniCalmanti2007, LucariniCalmanti2005} where $T_{\mathrm{fall}}$ is always very large. 

\begin{figure}
\centering
    	\subcaptionbox{} [0.45\textwidth]
    {\includegraphics[width=0.45\textwidth]{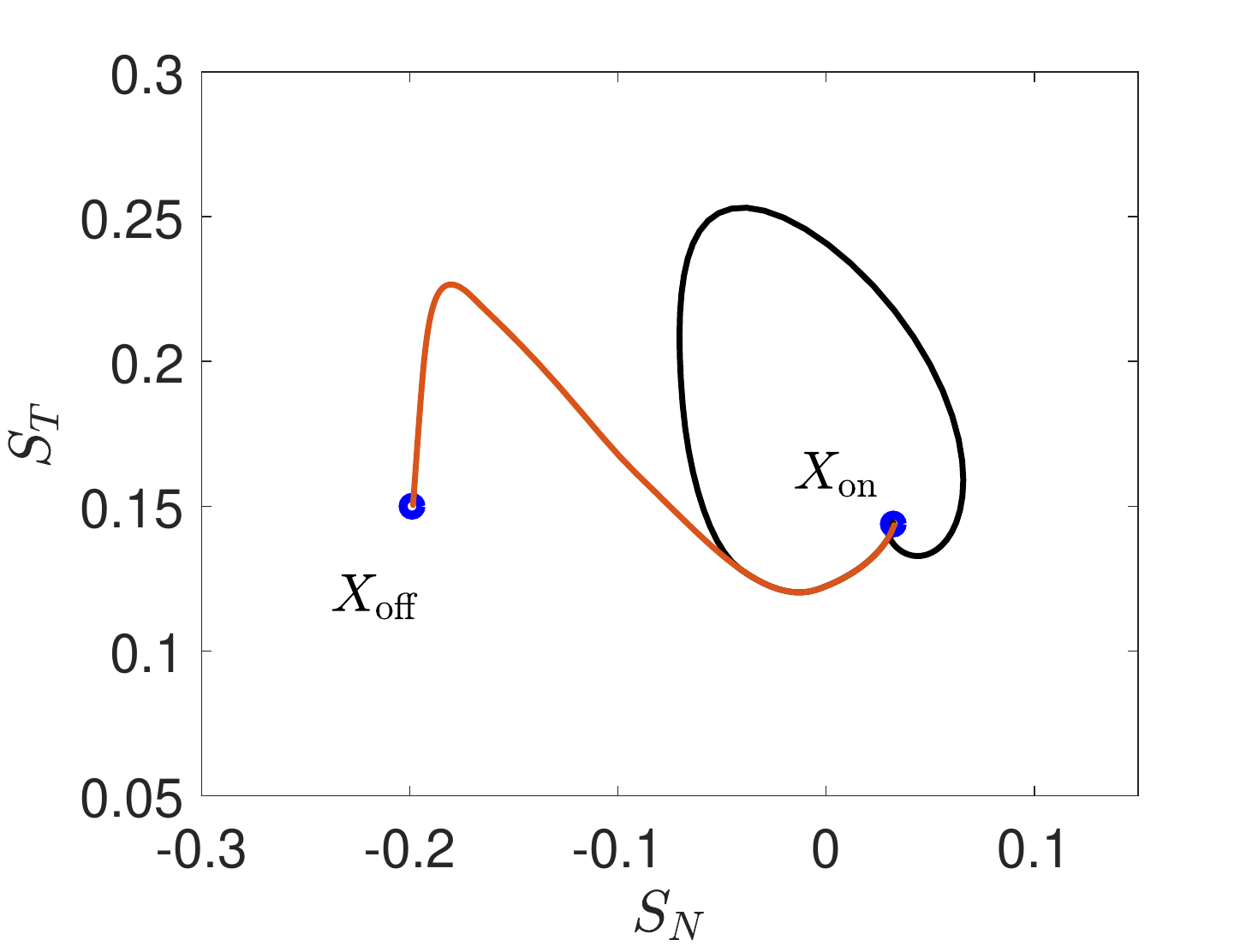}}
    	\subcaptionbox{} [0.45\textwidth]
    {\includegraphics[width=0.45\textwidth]{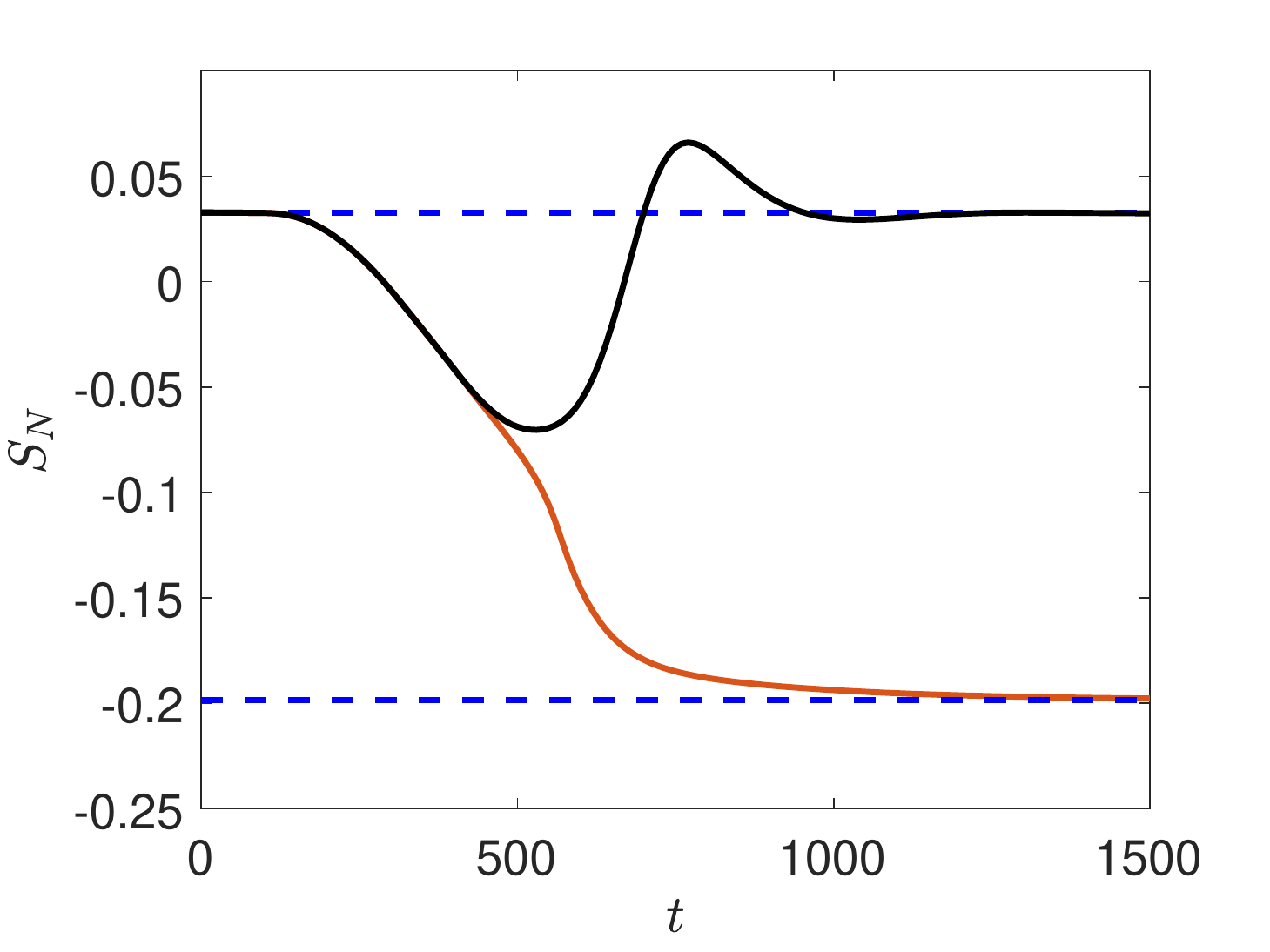}}
 \caption{Trajectories of (\ref{eq:AOMC_model_q>0_3box},\ref{eq:AOMC_model_q<0_3box}) for $2\times \mathrm{CO}_2$ parameters and piecewise linear time-dependent parameter $H(t)$ defined by \eqref{eq:H_PWL}, the parameter values are $T_{\mathrm{pert}} = 200$, $H_0=0$,  $H_{\mathrm{pert}} = 0.5$, $T_{\mathrm{rise}} = 200, T_{\mathrm{fall}} = 100$ for the black trajectory, and $T_{\mathrm{fall}} = 200$, $T_{\mathrm{rise}} = 200$  for the dark red trajectory. (a) shows the phase portrait and (b) the time profile of $S_N$ (for time profile of $S_T$ see Figure~\ref{fig:Btip}). In all cases we start at an initial condition on $X_{\mathrm{on}}$.} 
 
 \label{fig:timeseries_HPWL}
\end{figure}

As an example of the sort of critical transition that can arise, Figure~\ref{fig:timeseries_HPWL} shows the effect of two similar PWL perturbations that differ only in $T_{\mathrm{fall}}$. As can be seen in the phase plane (a) and time series (b) these lead to quite different outcomes at the end of the perturbation.

\subsection{Resilience time: B-tipping and avoided B-tipping for instantaneous forcing changes}
\label{sec:Btipinst}

\begin{figure}
\centering
\includegraphics[width=0.8\textwidth]{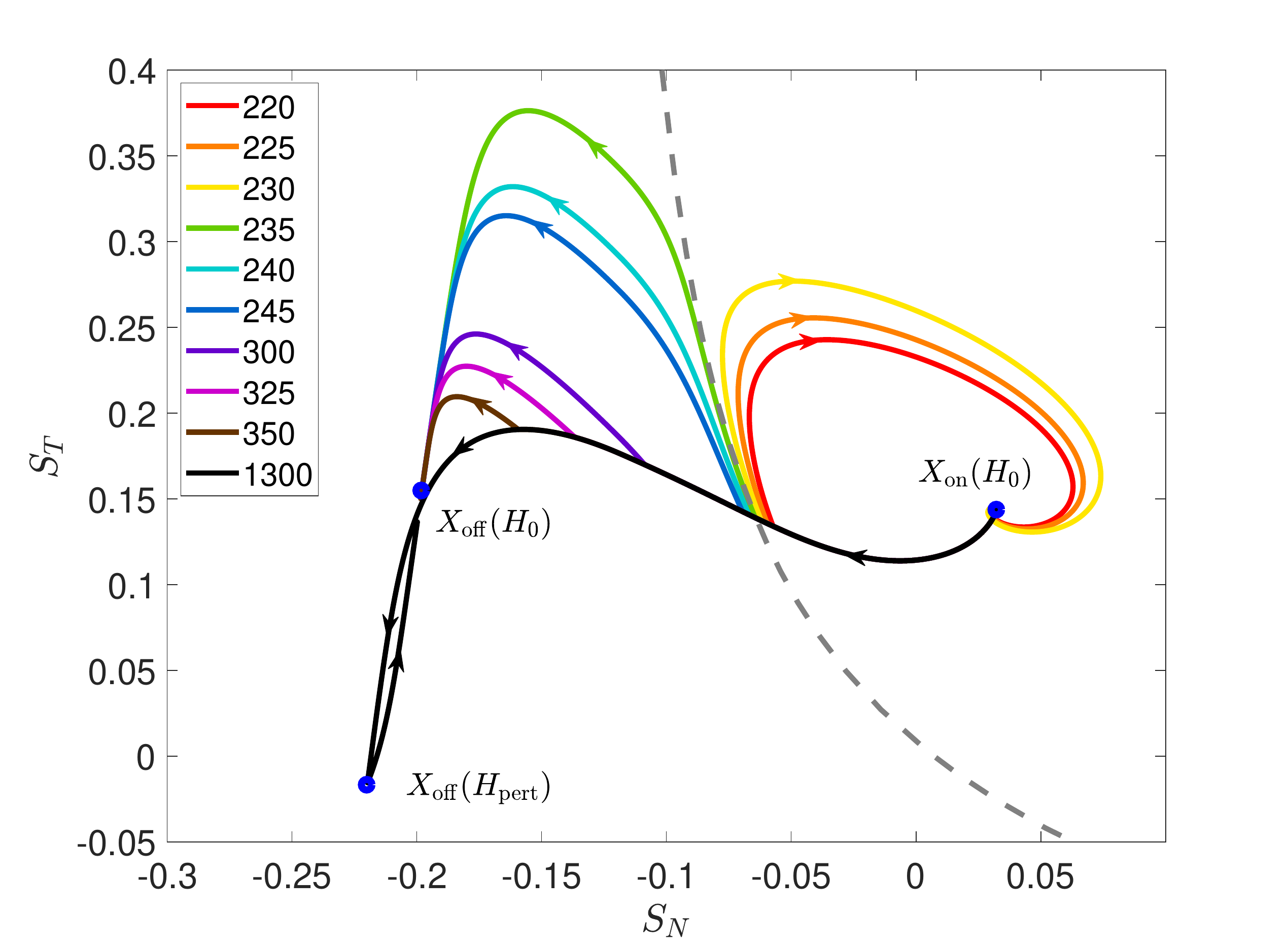}
 
\caption{Trajectories of (\ref{eq:AOMC_model_q>0_3box},\ref{eq:AOMC_model_q<0_3box}) ($2\times \mathrm{CO}_2$ parameters) with $H$ defined by (\ref{eq:H_PWL}) for the ``press'' perturbation with $\Hp=0.5$ and $T_{\mathrm{rise}}=T_{\mathrm{fall}}=0$.  The initial hosing value is fixed at $H_0=0$, and $T_{\mathrm{pert}}$ (in years) is indicated by the line colour. An initial condition is taken close to $X_{\mathrm{on}}$ for $H=H_0$. Note that longer durations result in switching to $X_{\mathrm{off}}$, and there is a critical duration, $234$ years, that gives a connection to the saddle state $X_{\mathrm{saddle}}$ (not shown). The grey dashed line represents the stable manifold of the saddle equilibrium (off the top of the figure) for $H=0$. 
 }
 \label{fig:pp_Hpress}
\end{figure}

A special case of the PWL perturbation (\ref{eq:H_PWL}) is when $T_{\mathrm{rise}}=T_{\mathrm{fall}}=0$: an instantaneous change from $H_0$ to the perturbed $H_{\mathrm{pert}}$ and then back. In all cases we start from the $X_{\mathrm{on}}$ state for $H=0$. This type of perturbation has been studied extensively in ecological settings \cite{ratajczak2017} where it is called a ``press''. We explore the behaviour of this case when varying $H_{\mathrm{pert}}$ and $T_{\mathrm{pert}}$, fixing $H_0=0$: an example is depicted in Figure \ref{fig:pp_Hpress}. The black line shows the initial departure from the equilibrium state for $H=H_0$.  All trajectories follow this path in the $H=\Hp$ phase plane for the time $\Tp$ corresponding to the line colour.  When $t>\Tp$ trajectories then make an excursion back to one of the two stable equilibria $X_{\mathrm{on/off}}$ for $H=H_0$. If a trajectory returns to $X_{\mathrm{on}}$ we say it does not tip, while if it approaches $X_{\mathrm{off}}$ then we say it tips.

\begin{figure}
\centering
    {\includegraphics[width=0.6\textwidth]{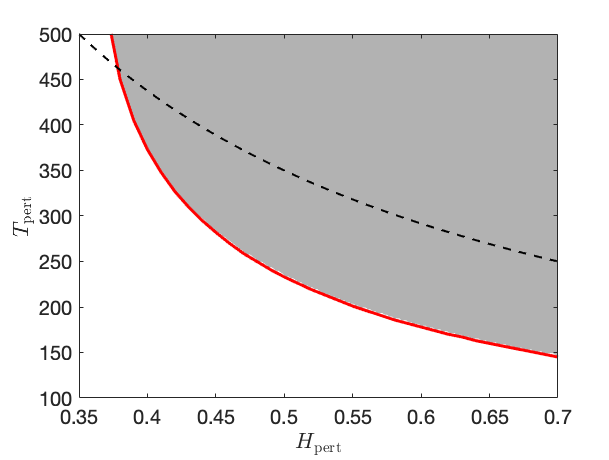}}
    
 \caption{
Final state of (\ref{eq:AOMC_model_q>0_3box},\ref{eq:AOMC_model_q<0_3box}) starting at $X_{\mathrm{on}}$ for $H=0$ subjected to the perturbation $H_{\mathrm{pwl}}$ for a range of $T_{\mathrm{pert}}$ and $H_{\mathrm{pert}}$, for the ``press'' perturbation with $T_{\mathrm{rise}}=T_{\mathrm{fall}}=0$ ($2\times \mathrm{CO}_2$ parameters). White indicates eventual return to $X_{\mathrm{on}}$ (not tipping), while grey indicates tipping to $X_{\mathrm{off}}$ after the perturbation. The red line indicates the threshold where there is a source-saddle connection induced by the perturbation (``resilience time'' for a given perturbation strength).  The black dashed line represents a constant perturbation volume, \emph{i.e.} $H_{\mathrm{pert}}\times T_{\mathrm{pert}}=175$. 
}
\label{fig:Hpert_Tcrit}
\end{figure}

The critical situations that divide these ``tipping'' and ``non-tipping'' responses can be estimated as follows. Comparing Figures~\ref{fig:pp_Hpress}a and \ref{fig:threebox_basins_H} we note that the divergence corresponds to where the black trajectory intersects with the basin boundary for $H=0$ (also shown on Figure 10). This critical situation corresponds precisely to the situation where a trajectory starting at $X_{\mathrm{on}}$ is on the stable manifold of the saddle $X_{\mathrm{saddle}}$ when the perturbation turns off. This corresponds to there being a ``sink-saddle connection'' for the perturbation we consider. We illustrate this critical value (``resilience time'') for different values of $H_{\mathrm{pert}}$ (mostly $> H_{\mathrm{Hopf}}$) in the case $T_{\mathrm{rise}}=T_{\mathrm{fall}}=0$ and $H_0=0$ in Figure~\ref{fig:Hpert_Tcrit}. Note that there is not an exact relationship between whether tipping occurs and the total amount of fresh water input $H_{\mathrm{pert}}\times T_{\mathrm{pert}}$. In some cases a short, intense pulse can cause tipping while a longer but weaker pulse containing the same total fresh water does not (compare red and black dashed curves in the Figure). Physically this is because for the long, weak pulse the circulation (including the gyre diffusion term $K_N(S_T - S_N)$) has longer to flush the extra fresh water away from the North Atlantic (N box) (see also \cite{JW17}).

We now consider some illustrative examples of the effect of more general perturbations of the form (\ref{eq:H_PWL}) starting at $H_0=0$ and the upper branch.

\subsection{Resilience time: B-tipping and avoided B-tipping for gradual forcing changes}
\label{sec:Btip}

For more general perturbations of the type $H_{\mathrm{pwl}}$ illustrated in Figure~\ref{fig:PWL} we highlight that not only the perturbation and duration spent there, but also the timescales of getting there and returning will influence whether tipping will occur. In particular, even if $H_{\mathrm{pert}}>H_{\mathrm{Hopf}}$  B-tipping may be avoided: there can be a temporary resilience of the system even after crossing a bifurcation point if the return is fast enough, as already seen in Section~\ref{sec:Btipinst}.

\begin{figure}
\centering
    	\subcaptionbox{} [0.4\linewidth]
    {\includegraphics[scale = 0.4]{./AMOC_Btip_SN.eps}}
    	\subcaptionbox{} [0.4\linewidth]
    {\includegraphics[scale = 0.4]{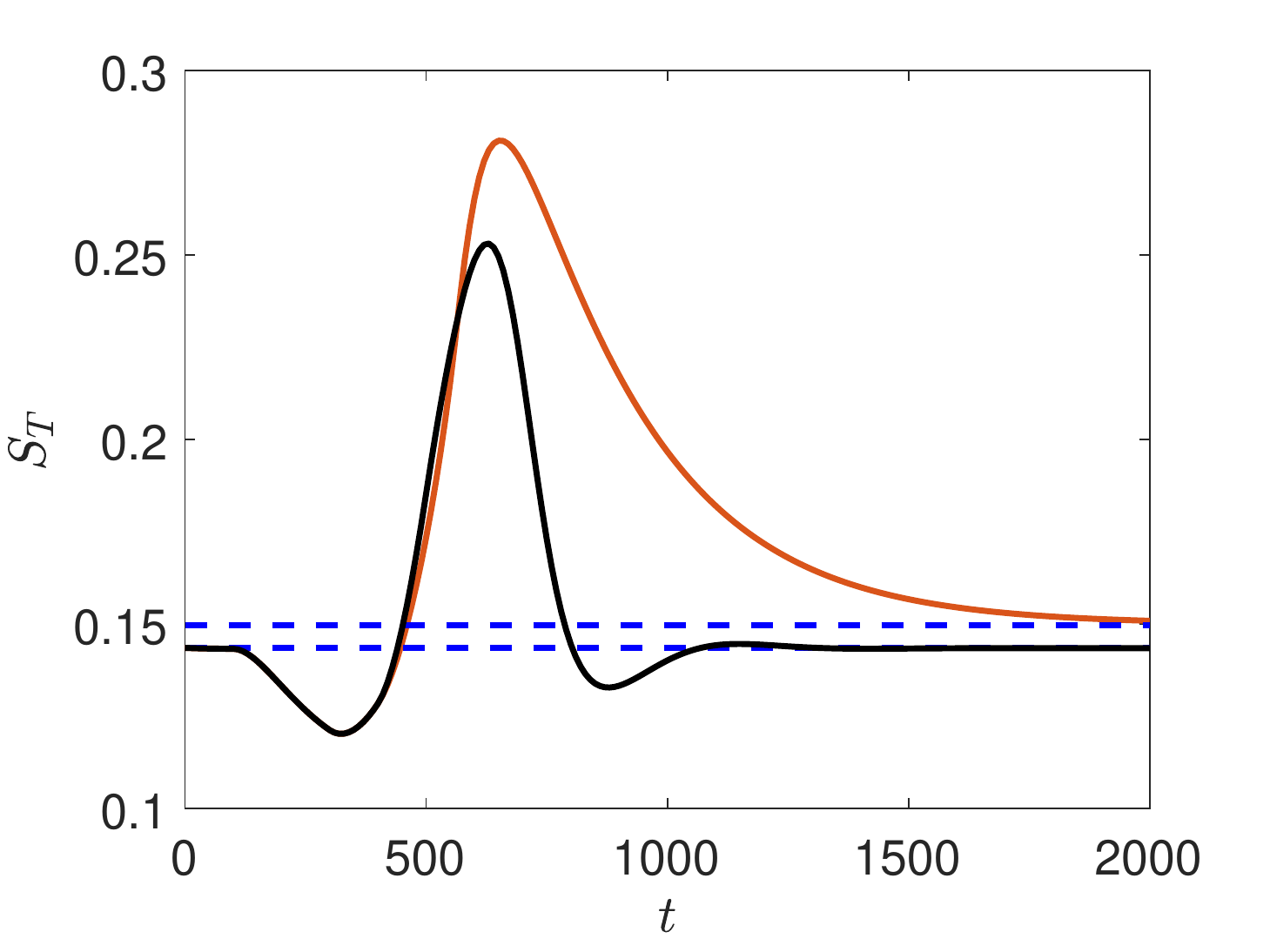}}
    	\subcaptionbox{} [0.4\linewidth]
    {\includegraphics[scale = 0.4]{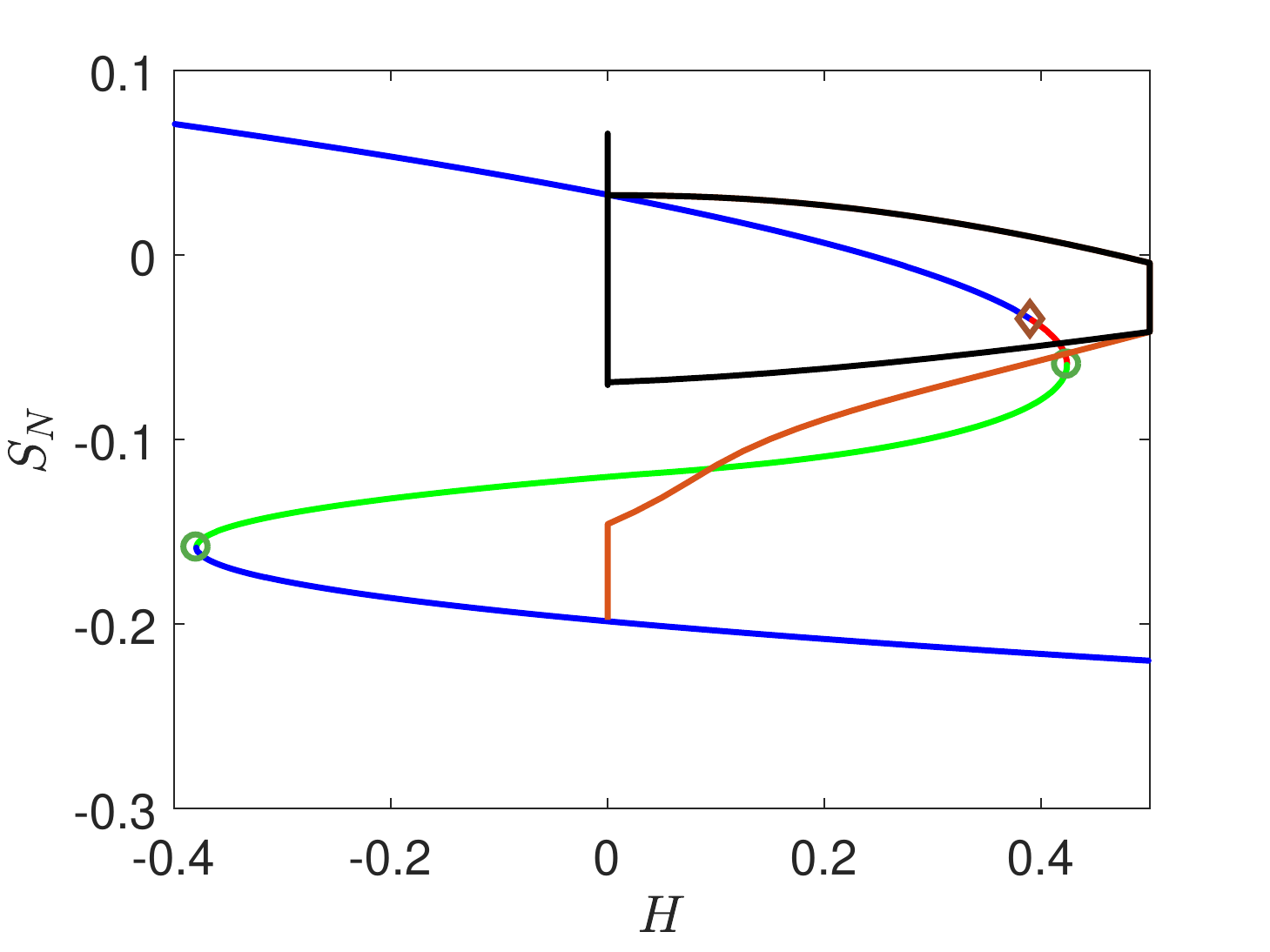}}
    	\subcaptionbox{} [0.4\linewidth]
    {\includegraphics[scale = 0.4]{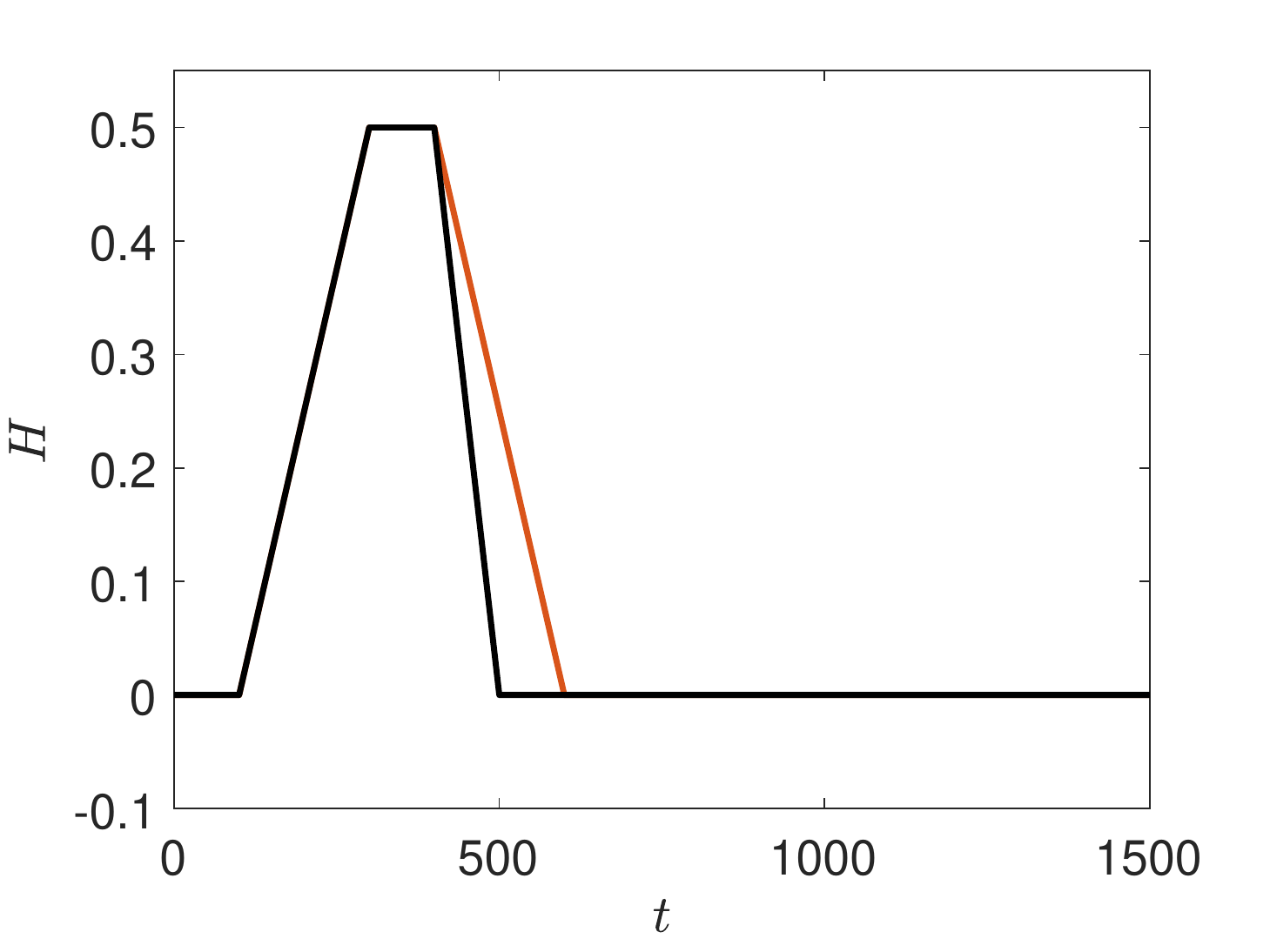}}
 \caption{Bifurcation-induced tipping for system (\ref{eq:AOMC_model_q>0_3box},\ref{eq:AOMC_model_q<0_3box}) for two trajectories: one tips (dark red) and the other does not (black) with $2\times \mathrm{CO}_2$. Figures (a) and (b) show the time series of $S_N$ and $S_T$ respectively. The dashed blue lines represent the two stable equilibria for $H=H_0$. Figure (c) shows the $S_N$ trajectory plotted over the bifurcation diagram. Figure (d) shows the piecewise linear forcing (\ref{eq:H_PWL}) against time with $H_0=0$, $H_{\mathrm{pert}} = 0.5$ and $T_0 = 100$, $T_{\mathrm{pert}} = 200$. The other parameter values are:  $T_{\mathrm{rise}} = T_{\mathrm{fall}} = 200$ for the dark red trajectory and $T_{\mathrm{rise}} = 200$, $T_{\mathrm{fall}} = 100$ for the black trajectory. Note that the parameter values of this figure are the same as in Figure~\ref{fig:timeseries_HPWL}. {\em Supplementary Movies {\tt MOC\_phas\_Btip.mp4} and {\tt MOC\_phas\_Btrack.mp4} show animations corresponding to the red and black trajectories respectively.}}
 
 \label{fig:Btip}
\end{figure}

In the case that $H_0<H_{\mathrm{Hopf}}<H_{\mathrm{pert}}$, the bifurcation diagram in Figure~\ref{fig:threeboxbifs2CO2} for $2\times \mathrm{CO}_2$ parameters shows that the lower branch is the unique attractor for $H=H_{\mathrm{pert}}$. This means that regardless of $T_{\mathrm{rise}}$ and $T_{\mathrm{fall}}$ we expect to see tipping onto the lower branch if $T_{\mathrm{pert}}$ is large enough. However, if some combination of $T_{\mathrm{rise}}$, $T_{\mathrm{fall}}$ and $T_{\mathrm{pert}}$ is sufficiently small we may avoid B-tipping. Figure~\ref{fig:Btip} illustrates these two scenarios, the black trajectory shows the avoiding tipping behaviour while the dark red trajectory shows B-tipping.   

\begin{figure}
\centering
\includegraphics[width=0.32\textwidth]{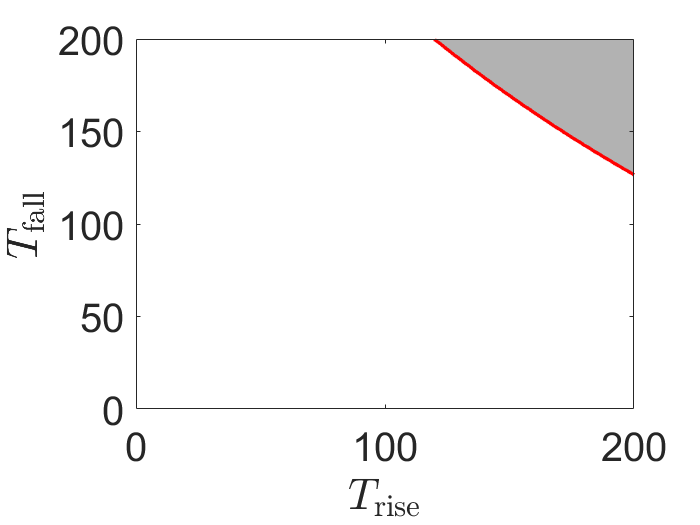}~
\includegraphics[width=0.32\textwidth]{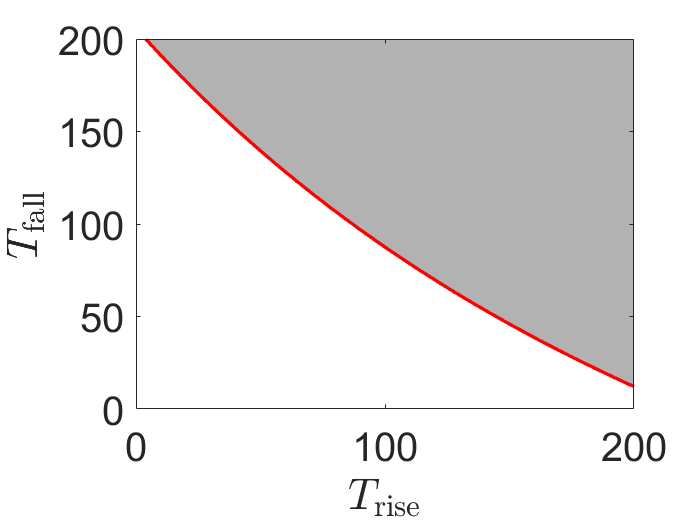}~
\includegraphics[width=0.32\textwidth]{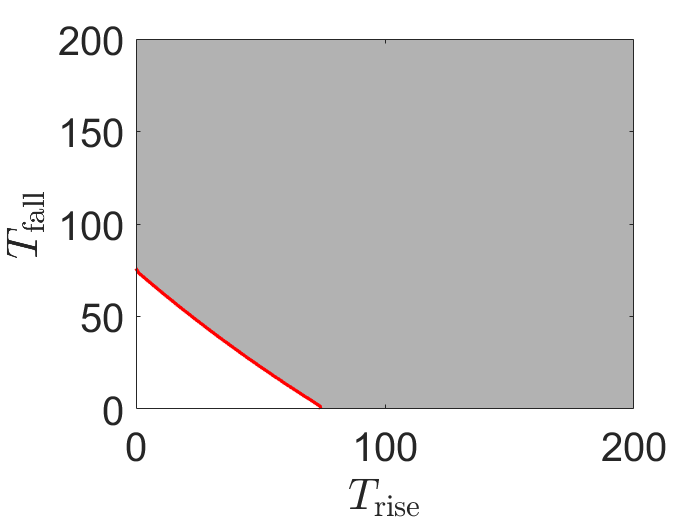}
\caption{Results of applying the piecewise linear perturbation $H_{\mathrm{pwl}}(t)$ starting from $X_{\mathrm{on}}$ with $H_{\mathrm{pert}}$ beyond the Hopf bifurcation, with $H_0=0$, $H_{\mathrm{pert}}=0.5$, $T_{\mathrm{pert}}=100$ (left), $150$ (middle), $200$ (right) and $2\times \mathrm{CO}_2$ parameters. Otherwise as in Figure~\ref{fig:Hpert_Tcrit}:
white indicates an eventual return to $X_{\mathrm{on}}$ (does not tip), while grey ends at the $X_{\mathrm{off}}$ state. The threshold is indicated in red. Observe that rapid enough rates (white regions of figures) gives temporary resilience, i.e. the system avoids tipping in spite of crossing the threshold. 
}
\label{fig:tip_after_Hopf}
\end{figure}

For fixed $T_{\mathrm{pert}}$ and $H_{\mathrm{pert}}=0.5$ we characterise in Figure~\ref{fig:tip_after_Hopf} the regions of $(T_{\mathrm{rise}}, T_{\mathrm{fall}})$ where we get tipping onto the lower branch as a result of the perturbation. For large enough $T_{\mathrm{pert}}$ we find only tipping. However for $T_{\mathrm{pert}}<235$ we find that a rapid enough combination of rise and fall can avoid tipping. Again, whether tipping occurs is not solely determined by the total amount of fresh water added, with shorter, more intense perturbations being more effective in inducing tipping.

\subsection{R-tipping, avoided R-tipping and basin boundaries}
\label{sec:Rtip}

\begin{figure}
\centering
    	\subcaptionbox{} [0.4\linewidth]
    {\includegraphics[scale = 0.4]{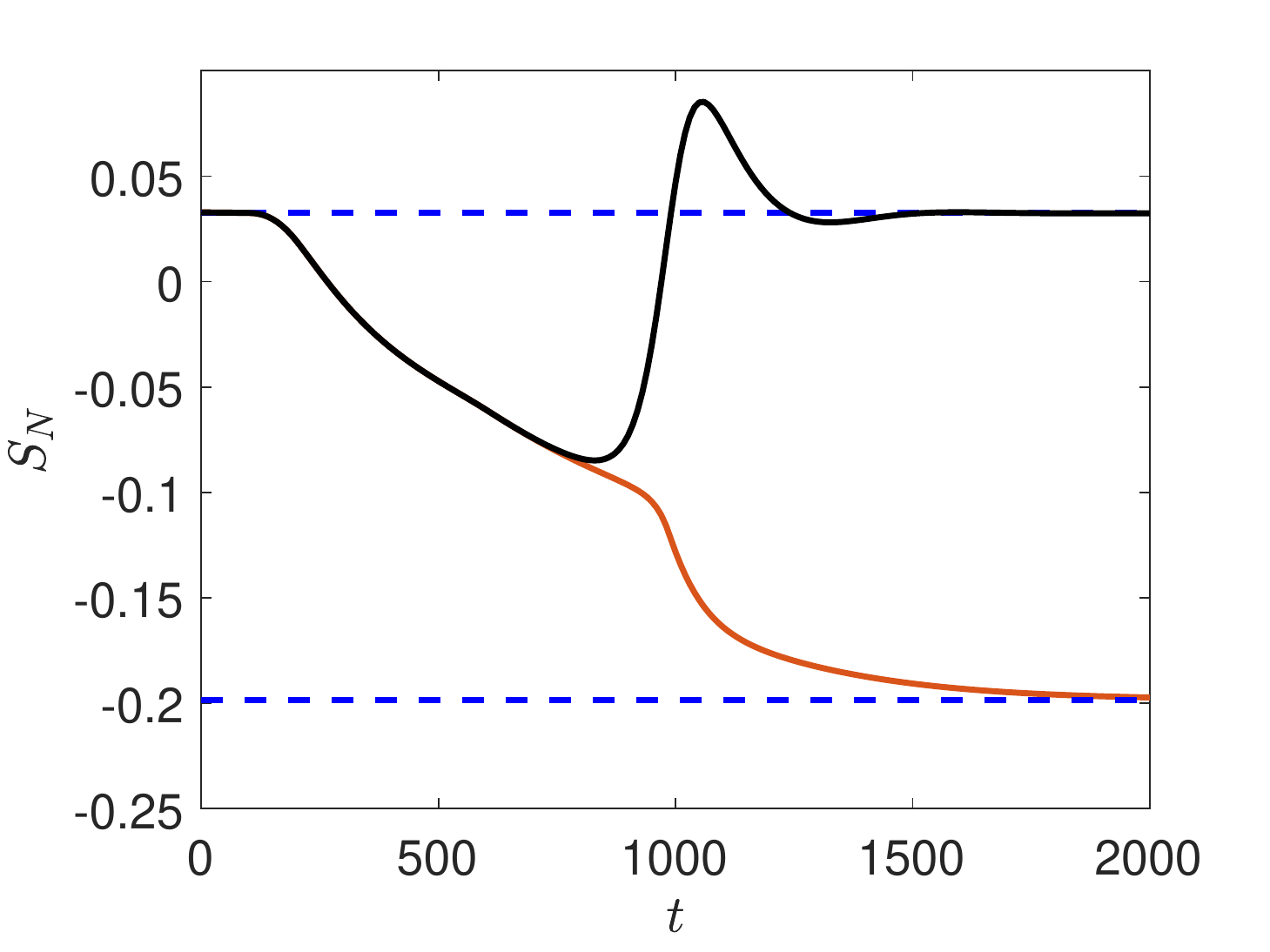}}
    	\subcaptionbox{} [0.4\linewidth]
    {\includegraphics[scale = 0.4]{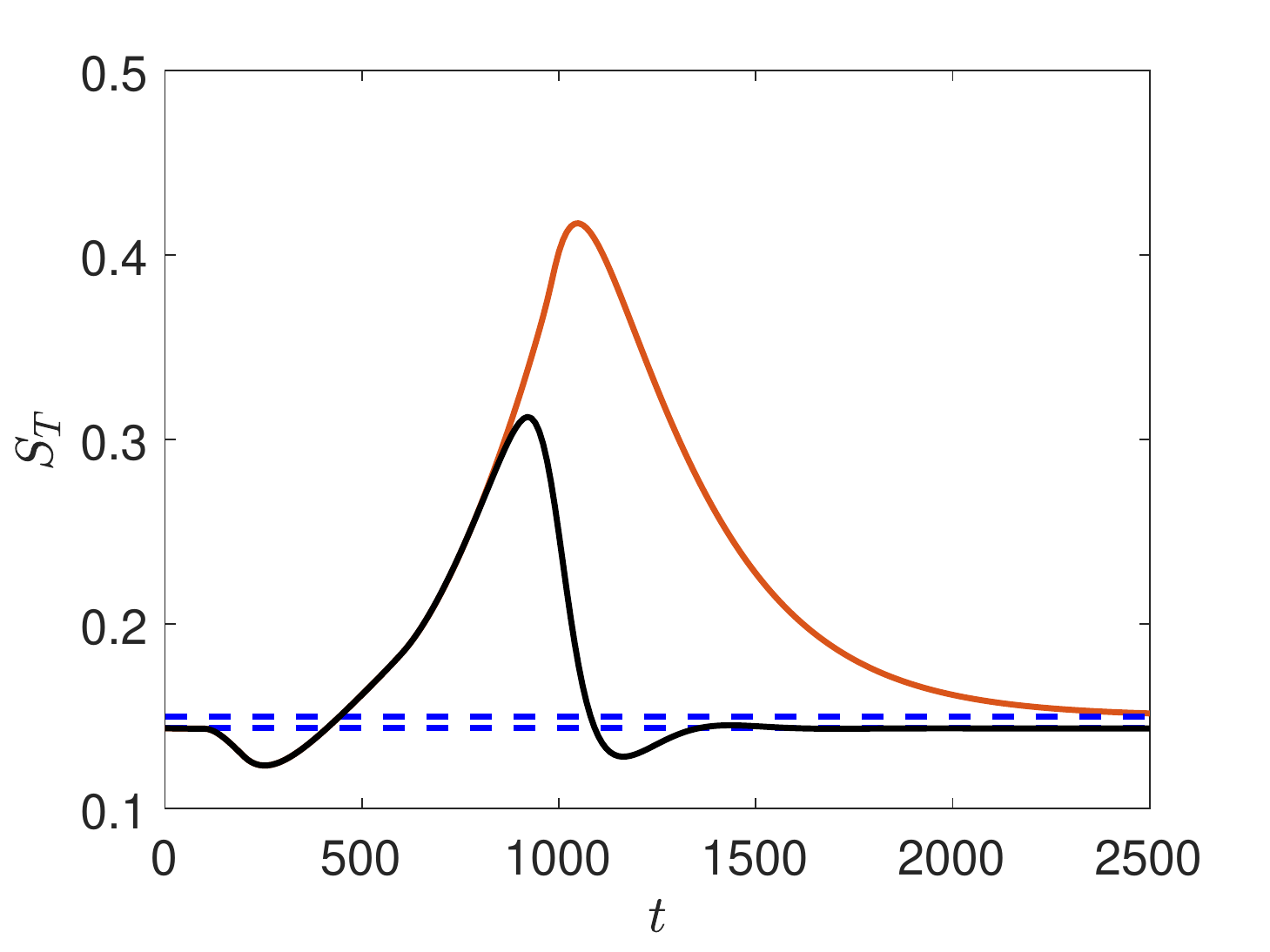}}
    	\subcaptionbox{} [0.4\linewidth]
    {\includegraphics[scale = 0.4]{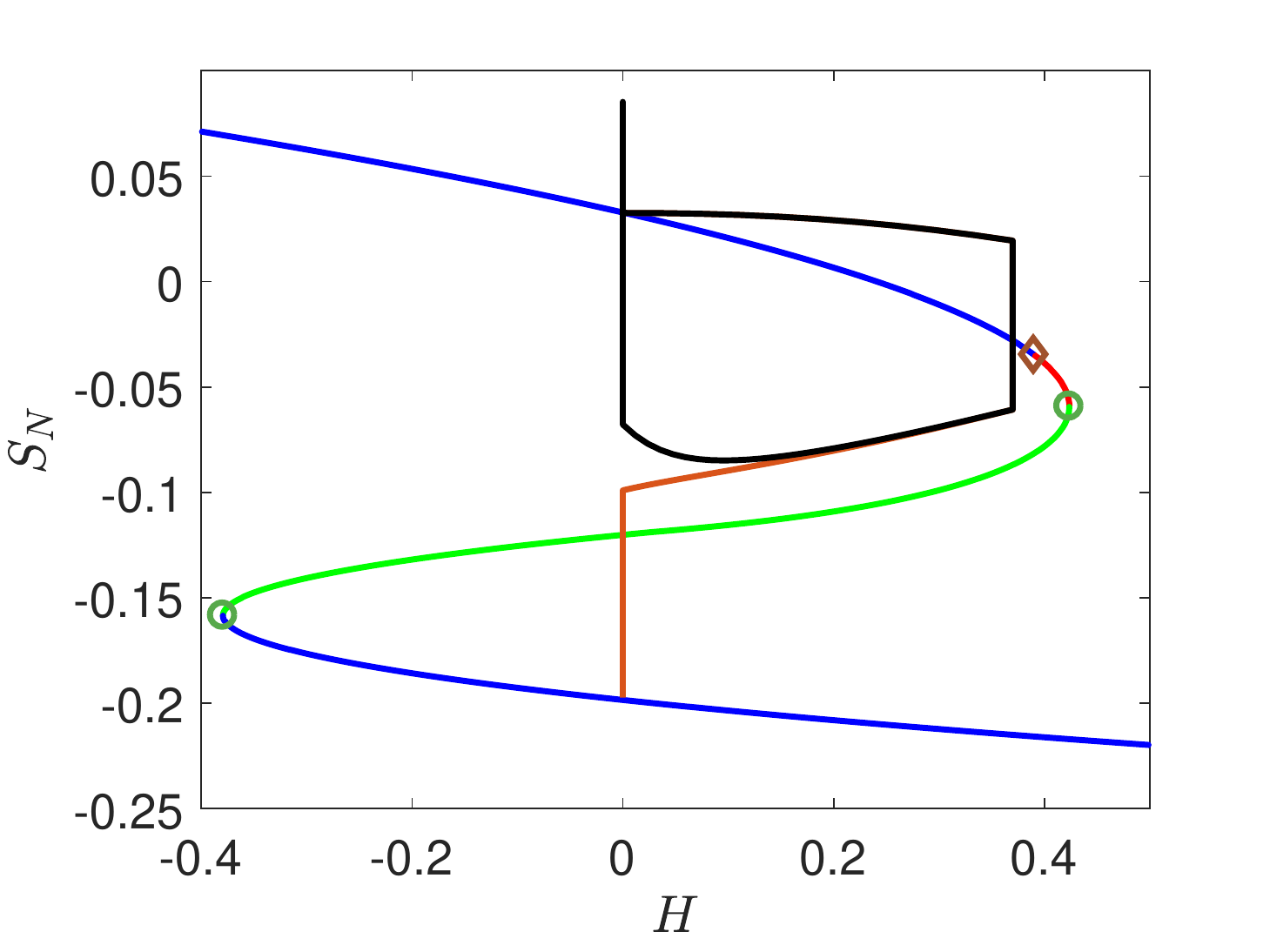}}
    	\subcaptionbox{} [0.4\linewidth]
    {\includegraphics[scale = 0.4]{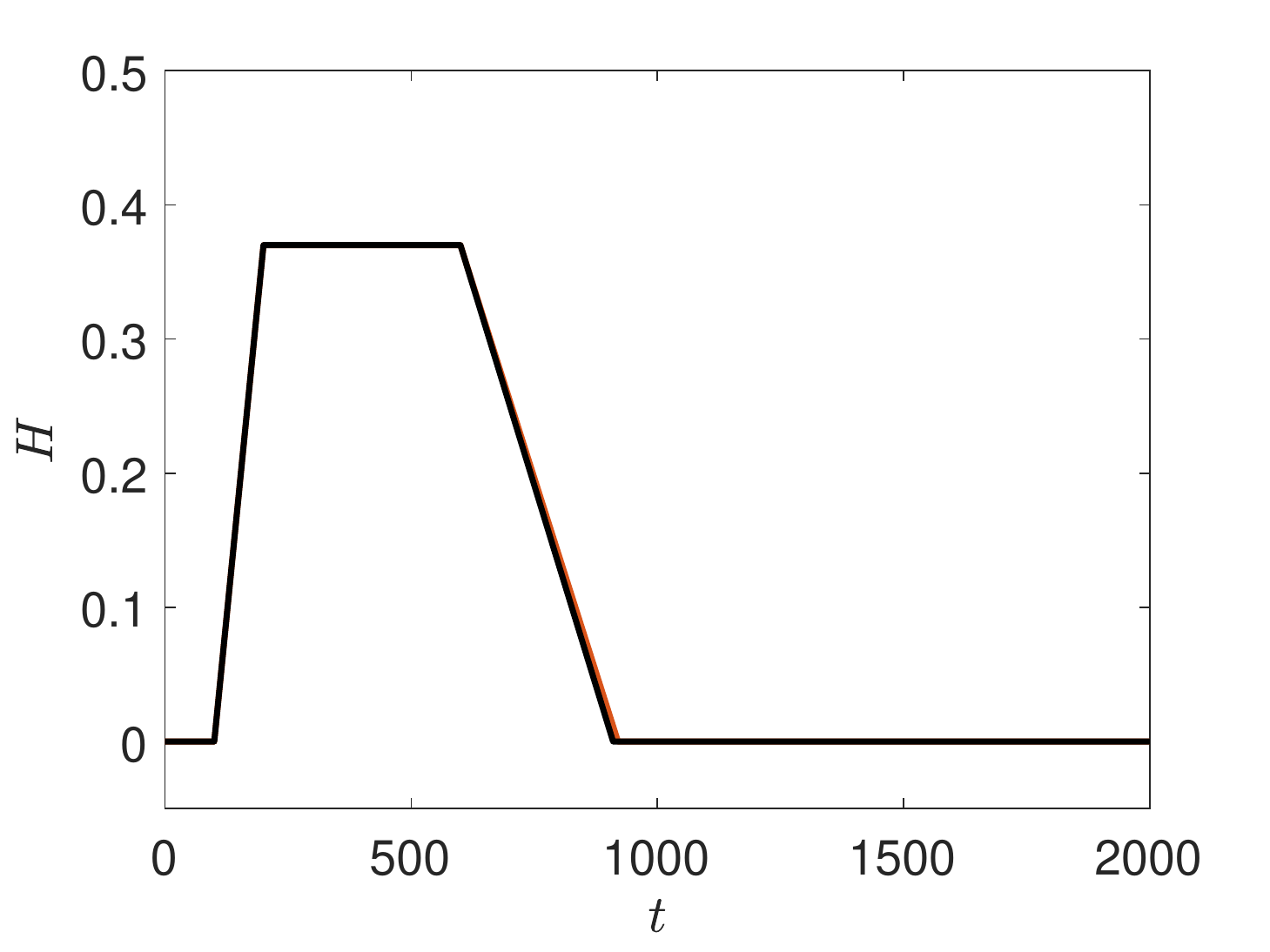}}
 \caption{Rate-induced tipping for system (\ref{eq:AOMC_model_q>0_3box},\ref{eq:AOMC_model_q<0_3box}) with $2\times \mathrm{CO}_2$ parameters. Two trajectories are shown, where one tips (dark red) and the other does not (black). (a) and (b) show the time series of $S_N$ and $S_T$ respectively the dashed blue lines represent the equilibria $X_{\mathrm{on/off}}$. (c) shows $S_N$ plotted over the bifurcation diagram. Finally, (d) shows the piecewise linear forcing against time with $H_0=0$ and $H_{\mathrm{pert}} = 0.37$ which is less than the $H_{\mathrm{Hopf}}$ where the upper branch loses its stability. The other parameter values are: $T_{\mathrm{rise}} = 100, \ T_{\mathrm{fall}} = 320, \ T_0 = 100$ and $T_{\mathrm{pert}} = 400$ for the dark red trajectory. $T_{\mathrm{rise}} = 100, \ T_{\mathrm{fall}} = 310, \ T_0 = 100$ and $T_{\mathrm{pert}} = 400$ for the black trajectory. {\em Supplementary Movies {\tt MOC\_phas\_Rtip.mp4} and {\tt MOC\_phas\_Rtrack.mp4} show animations corresponding to the red and black trajectories respectively.} }
 
 \label{fig:Rtip}
\end{figure}

The bifurcation diagram Figure~\ref{fig:threeboxbifs2CO2} might seem to suggest that a perturbation from $H_0$ to $H_{\mathrm{pert}}<H_{\mathrm{Hopf}}$ will not result in tipping to the lower branch for any choice of $T_{\mathrm{rise}},T_{\mathrm{fall}}$ and $T_{\mathrm{pert}}$. Indeed, Figure~\ref{fig:Rtip} shows the effect of a perturbation that increases and then decreases $H$ fast enough for there to be no tipping from the upper branch. On the other hand, for a less rapid decrease (the dark red trajectory in Figure~\ref{fig:Rtip}) we can find situations that switch to the lower branch.

\begin{figure}
\centering
\includegraphics[width=0.32\textwidth]{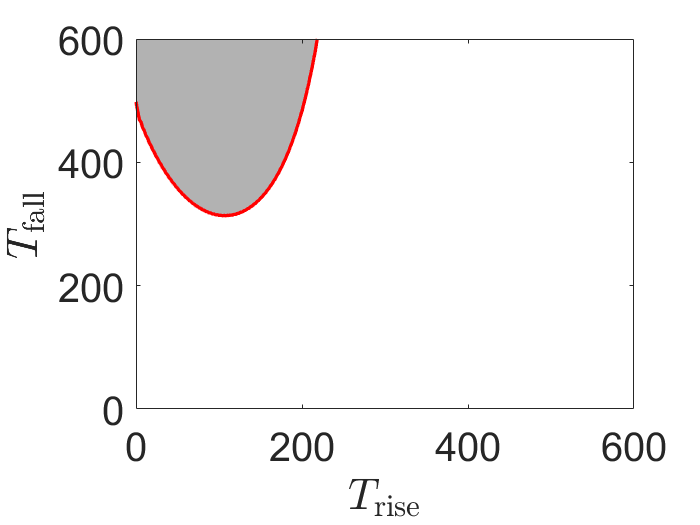}~
\includegraphics[width=0.32\textwidth]{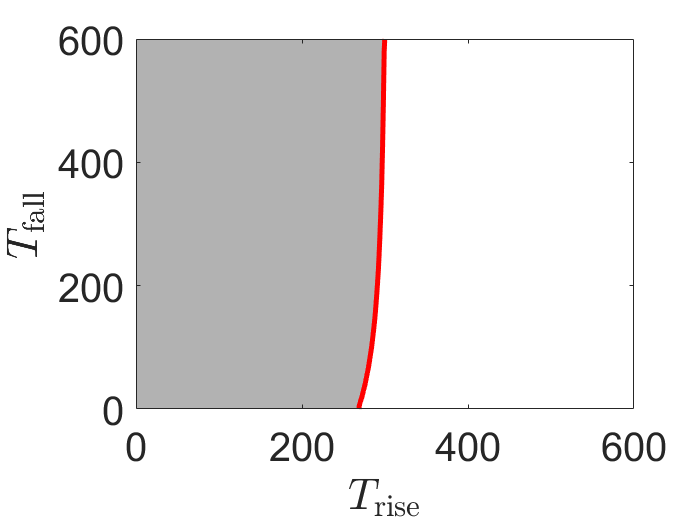}~
\includegraphics[width=0.32\textwidth]{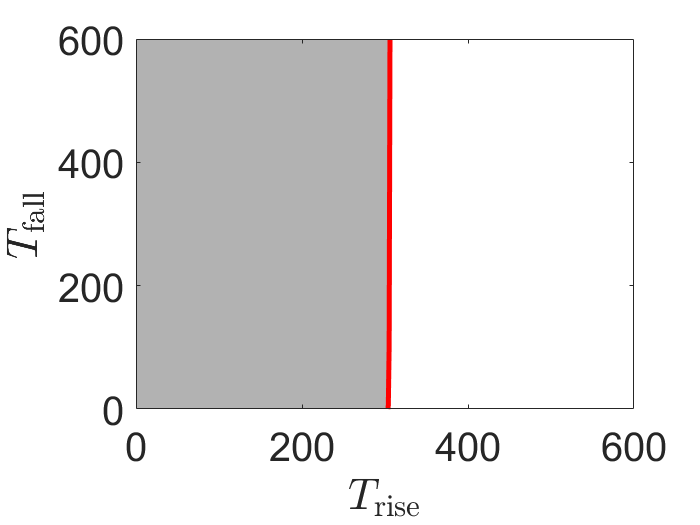}
\caption{
Eventual behaviour of the system as in Figure~\ref{fig:Hpert_Tcrit} for $H_{\mathrm{pwl}}$ with $H_{\mathrm{pert}}$ before the homoclinic, $H_0=0$, $H_{\mathrm{pert}}=0.37$, $T_{\mathrm{pert}}=400$ (left), $600$ (middle), $800$ (right), and $2\times \mathrm{CO}_2$ parameters. As before, white: does not tip, grey: tips, red: threshold. Note that too rapid a rise of the perturbation can give rate-induced tipping, even when the perturbation is held for a long time and the return is very slow.}
\label{fig:tip_before_hom}
\end{figure}

For fixed $T_{\mathrm{pert}}$ and $H_{\mathrm{pert}}=0.37$ we characterise in Figure~\ref{fig:tip_before_hom} the regions of $(T_{\mathrm{rise}}, T_{\mathrm{fall}})$ where we get tipping onto the lower branch as a result of the perturbation. For large enough $T_{\mathrm{pert}}$ we find tipping if $T_{\mathrm{rise}}$ is small enough, independent of $T_{\mathrm{fall}}$. 

We can interpret this behaviour as follows: for $T_{\mathrm{pert}}$ large enough, tipping depends on whether we are in the basin of attraction of the lower branch at the end of the rise. Because the basin for $H=0.125$ is bounded, a fast enough rise can leave the system outside the upper branch basin.

\section{Relationship to AMOC tipping in complex climate models}
\label{sec:gcm}

The analysis of Sections~\ref{sec:fivebox} and \ref{sec:tipping} provides insight into the bifurcation structure and tipping behaviour of the five box model and its three box reduction. Given that the box model has been shown to capture the dynamics of AMOC hysteresis and tipping in the FAMOUS AOGCM \cite{RWSH17}, it is natural to ask how relevant our insights are to the AMOC behaviour in complex climate models. For the present study we simply ask how the behaviour of the box model relates to existing GCM and EMIC studies. Future work will exploit the insights from the box model to target experimental design for further AOGCM studies, and hence make best use of limited computational resources.

A particular feature of the transient box model solutions can be seen in the ($S_N$,$S_T$) phase trajectories shown in idealised form in Figure~\ref{fig:threebox_sketchbifs}: trajectories in the vicinity of the upper branch (strong AMOC) equilibrium oscillate in a clockwise sense in phase space (whether they are approaching or diverging from the upper branch), but trajectories that are eventually destined for the lower branch (weak/reversed AMOC) eventually show anticlockwise curvature (or a sharp left turn at the middle branch), and approach the lower branch in a non-oscillatory way. This suggests a simple and in principle observable diagnostic in a more complex model (or the real world): anticlockwise trajectories in phase space suggest that the system is on the attractor of the lower branch. 

We use the above diagnostic to interpret the ``hosing hysteresis" experiments described by \cite{hawkins2011,RWSH17}. In these experiments hosing $H$ is slowly increased from $0$ to $1 \mathrm{Sv}$ over 2000 years, then decreased again. We know from \cite{RWSH17} that the five box model reproduces the dynamics of the FAMOUS AOGCM for these experiments, especially for the ``ramp up" phase where $H$ is increasing. The usual assumption in such experiments is that the forcing change is slow enough that the system remains close to its equilibrium state throughout. However, \cite{hawkins2011} noted that the AMOC appeared to remain ``on" for long after $H$ had passed the critical value where no ``on" equilibrium existed, suggesting significant transient resilience. Looking at Figure~\ref{fig:threebox_basins_H} we see that for states just outside the basin of the ``on'' state, the time to reach the ``off'' state equilibrium is of order hundreds to thousands of years; since this timescale coincides with the timescale of the forcing increase it is not surprising that the AMOC persists long after the forcing threshold $H_{\mathrm{usn}}$ has been passed.

\begin{figure}
\centering
    {\includegraphics[width=0.55\textwidth] {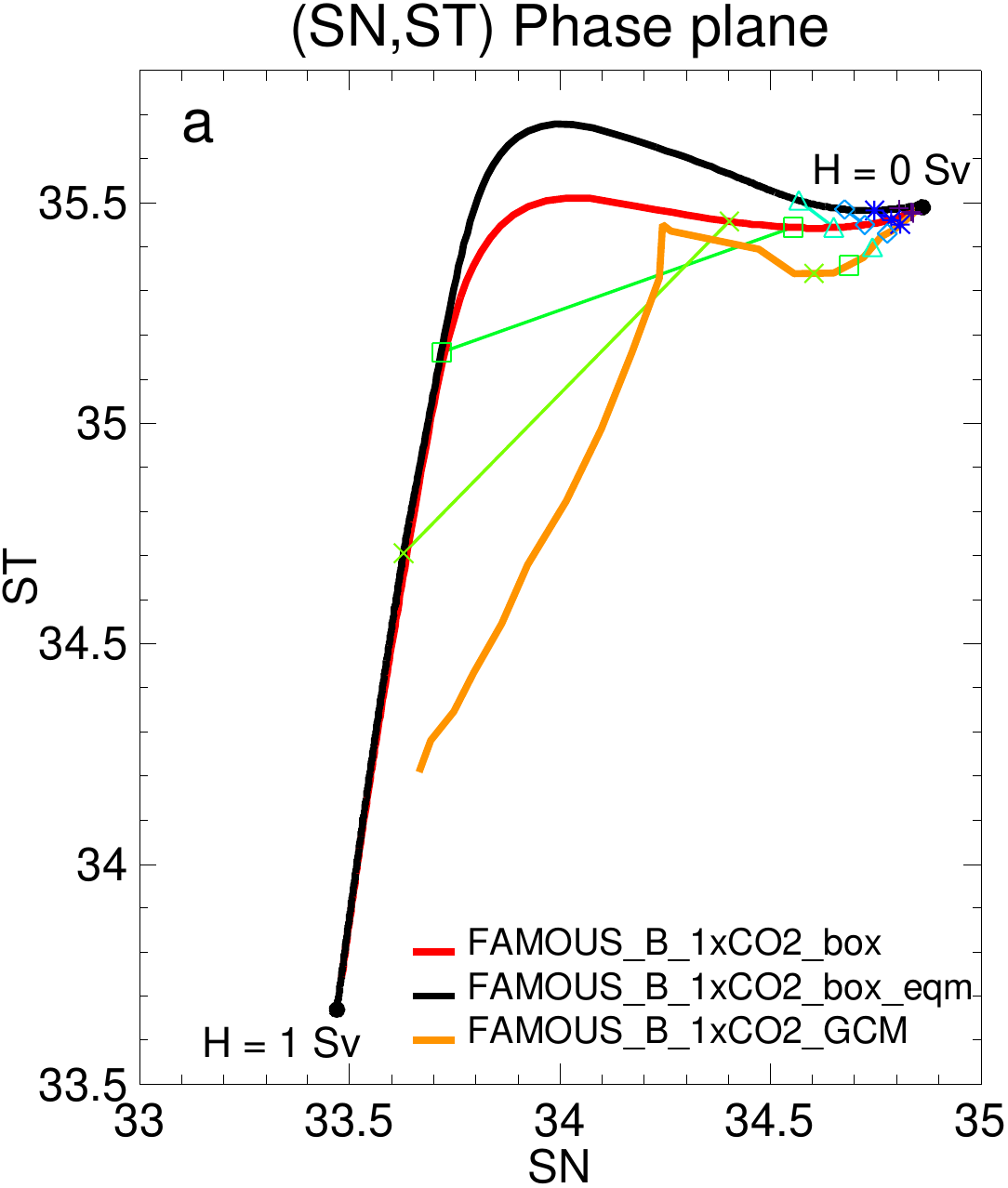}}
    {\includegraphics[width=0.7\textwidth] {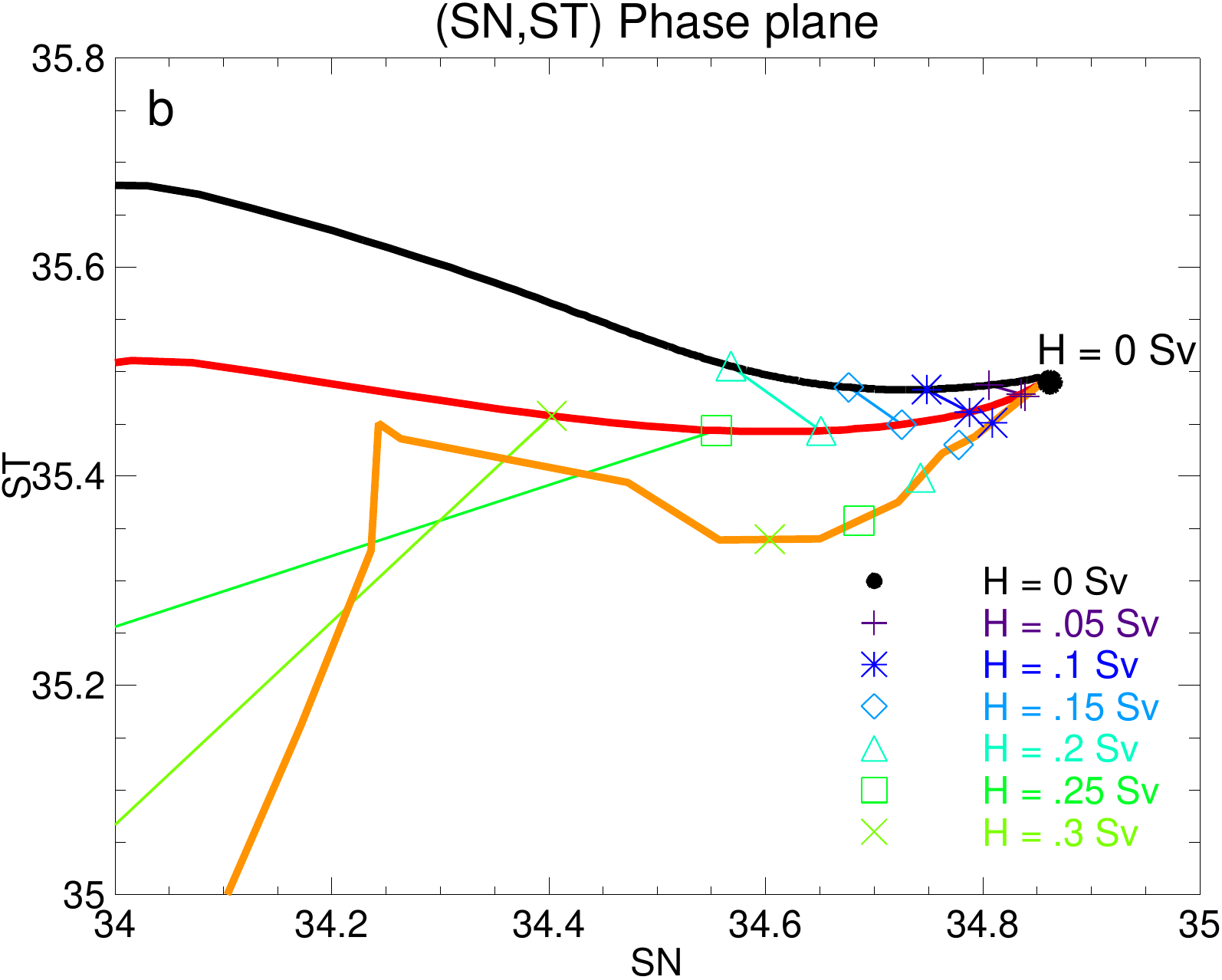}} 
     
 \caption{
 Trajectories in $(S_N,S_T)$ phase space for the ``ramp-up" phase of the hysteresis experiment ({H} slowly increasing from 0 to 1 Sv), for the $\rm{FAMOUS_B 1\times CO_2}$ AOGCM experiment of \cite{RWSH17} (orange) and for the corresponding five box model experiment (red). Symbols show the instantaneous state when the hosing $H$ reaches different values (intervals of 0.05 Sv hosing or 100 years in time). The black curve shows a repeat of the box model hysteresis experiment but with a 10 times slower timescale for increasing $H$, providing an estimate of the equilibrium state associated with that $H$ value. Transient states from the red trajectory are joined to the corresponding equilibrium states by lines. (a) Full ramp-up phase, (b) zoom in on early part (top right hand corner of (a)).
}
 
\label{fig:hysteresis_trajectories}
\end{figure}

\begin{figure}
\centering
    {\includegraphics[scale = 0.7] {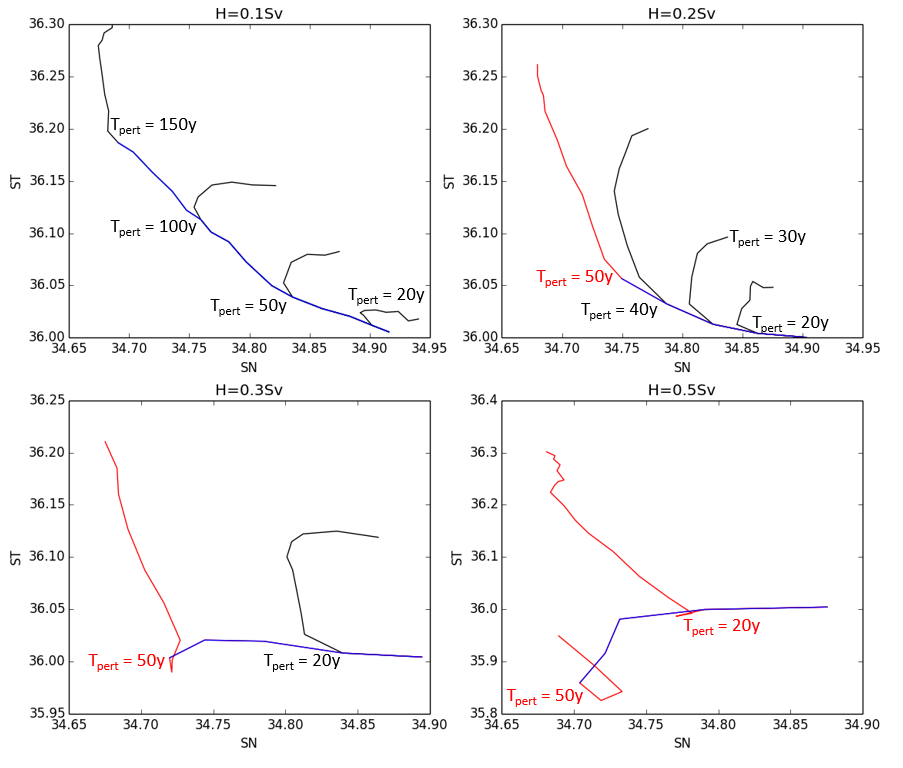}}
     
 \caption{
 Trajectories in $(S_N,S_T)$ phase space for the ``press'' experiments using the AOGCM HadGEM3-GC2 \cite{JW17}, with various values of $(H_{\mathrm{pert}},T_{\mathrm{pert}})$. The hosed phase of the run is shown in blue. Black trajectories show cases where the AMOC clearly recovers towards its initial sate after hosing stops, red trajectories show cases where the AMOC remains weak. In all the red cases, further integration of the AOGCM would be needed to determine the final state. 
}
 
\label{fig:GC2_press_traj}
\end{figure}

Figure~\ref{fig:hysteresis_trajectories} shows the $(S_N,S_T)$ phase plane for the ramp-up phase of the hosing hysteresis experiment in the five box model (red). Points on the trajectory are joined by blue/green lines to the equilibrium point corresponding to their instantaneous hosing value (estimated by re-running the hysteresis experiment with a ten times slower rate of increasing $H$, black, {\it{cf.}} \cite{LucariniCalmanti2005}). We see that initially the transient solution remains close to its corresponding upper branch equilibrium. However when $H$ increases beyond $H_{\mathrm{usn}}$ (around 0.22 Sv for this parameter setting, Table~\ref{tab:comparebifsH}) and the upper branch ceases to exist, the transient solution remains in a strong AMOC state (large $S_N$, $S_T$), even though its equilibrium is now a weak AMOC state (low $S_N$, $S_T$). This change occurs at a point in the trajectory where $S_T$ starts to increase, while $S_N$ continues to decrease, an indication that a positive salinity advection feedback has started to dominate between the $T$ and $N$ boxes, reducing the amount of salty water carried northwards and so causing further MOC weakening(\cite{LucariniCalmanti2005},\cite{RWSH17}). The trajectory for the corresponding FAMOUS experiment is qualitatively similar (orange curve): again $S_T$ starts to increase before the trajectory makes a sharp left turn towards the lower branch state. The change from clockwise to anticlockwise curvature indicates an approach to the lower branch state, but the critical value of hosing may have already been passed by the time the curvature changes; the earlier change from decreasing to increasing $S_T$ may be a better indicator of when $H_{\mathrm{usn}}$ has been passed, but this result may be scenario-dependent as it does not appear to apply for the ''press'' scenario (see Figures~\ref{fig:pp_Hpress}, \ref{fig:GC2_press_traj}).

\begin{figure}
\centering
    {\includegraphics[width=0.45\textwidth]{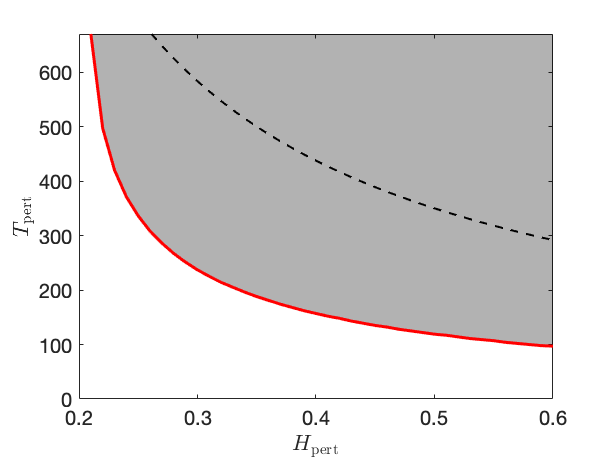}}
     {\includegraphics[width=0.45\textwidth]{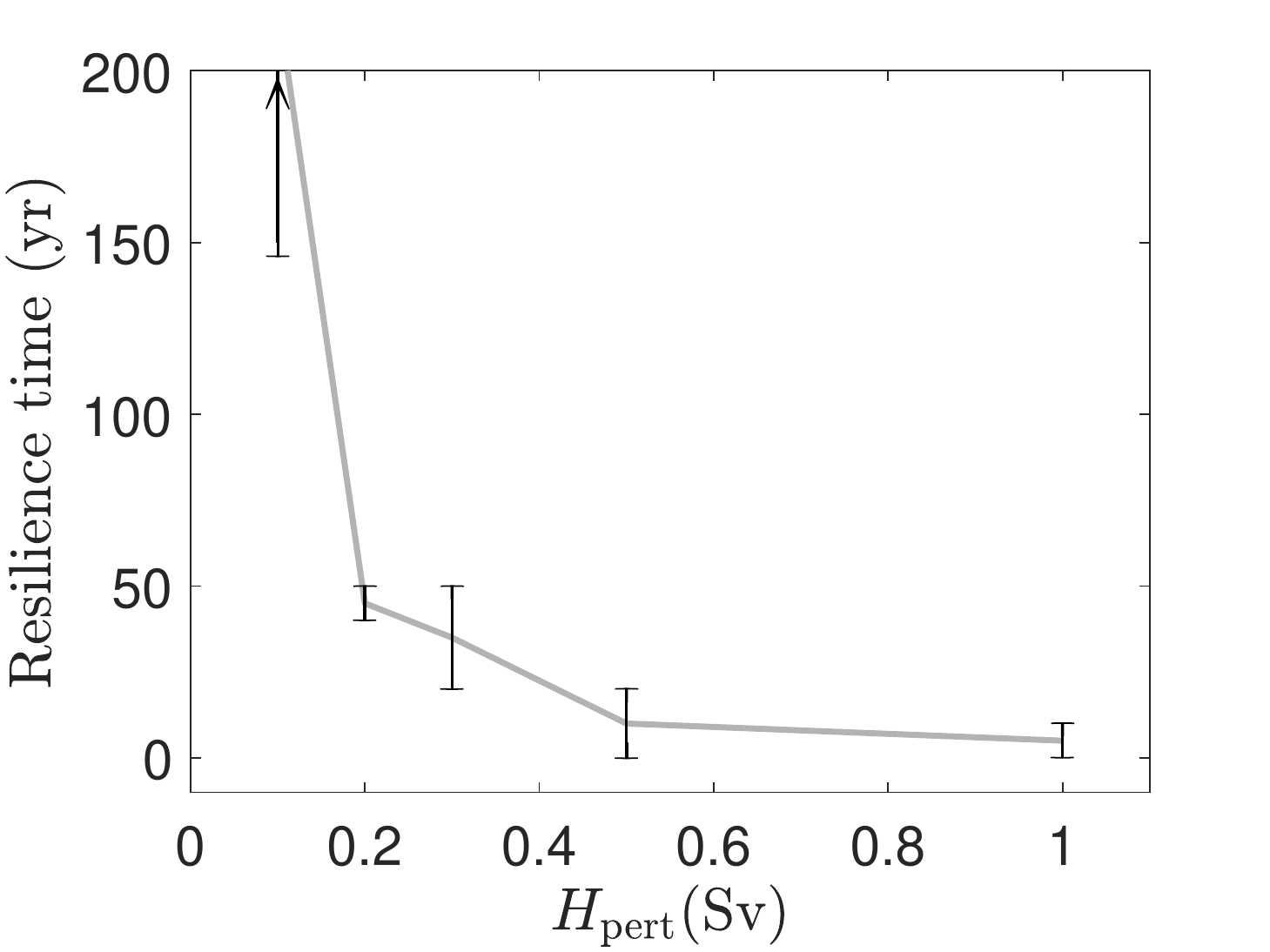}}
 \caption{
(a) Final state of (\ref{eq:AOMC_model_q>0_3box},\ref{eq:AOMC_model_q<0_3box}) starting at $X_{\mathrm{on}}$ subjected to the perturbation $H_{\mathrm{pwl}}$ for a range of $T_{\mathrm{pert}}$ and $H_{\mathrm{pert}}$, with $T_{\mathrm{rise}}=T_{\mathrm{fall}}=0$ ($1\times \mathrm{CO}_2$ FAMOUS parameters).  The white region indicates eventual return to $X_{\mathrm{on}}$ (not tipping), while grey indicates tipping to $X_{\mathrm{off}}$ after the perturbation. The red line shows the resilience time and  the black dashed line represents a constant integrated volume flux, \emph{i.e.} $H_{pert}\cdot T_{pert}=175$ 
b) The resilience time for HadGEM3-GC2 from \cite{JW17}. Bars indicate the range between the greatest $T_{\mathrm{pert}}$ for which there is a return to the initial state and the smallest $T_{\mathrm{pert}}$ for which there is possible tipping. The line joining the bars is for illustrative use only.}
 
\label{fig:res_time}
\end{figure}

We now consider the ``press" forcing scenario studied for the box model in Section~\ref{sec:Btipinst}. Jackson and Wood \cite{JW17} have recently studied resilience times in a CMIP6-generation AOGCM, HadGEM3-GC2 under such a scenario, for a range of values of $H_{\mathrm{pert}}$  and $T_{\mathrm{pert}}$. Figure  \ref{fig:res_time} shows the resilience times for the box model (with $1\times \mathrm{CO}_2$ parameters) and for HadGEM3-GC2. We do not expect quantitative comparison as HadGEM3-GC2 has different bifurcation points from FAMOUS (generally showing AMOC collapse at lower $H$ values). However, we do note that the forms of the two curves are the same, and the result that short, intense perturbations are more likely to cause tipping than long, weak perturbations is consistent across the models \cite{JW17}. Figure \ref{fig:GC2_press_traj} shows the phase space trajectories for the various HadGEM3-GC2 experiments of \cite{JW17}. The AOGCM trajectories compare qualitatively with those for the box model in Figure \ref{fig:pp_Hpress}: cases for which the AMOC clearly recovers after the forcing is removed show the same clockwise trajectories, while the cases where there is not a clear recovery (shown in red) mostly appear to be close to the dividing trajectory between recovery and collapse ({\it{cf.}} dashed line in  Figure \ref{fig:pp_Hpress}). In these cases the AOGCM AMOC appears to be in an intermediate state and it was not clear to \cite{JW17} whether that state would persist, revert to a strong AMOC or collapse (see Figure 1 of \cite{JW17}). Comparison to Figure \ref{fig:pp_Hpress} with a press scenario suggests that the AOGCM trajectories are not inconsistent with the box model but that the runs would need to be continued for longer to resolve this.

Overall the box model dynamics appears to provide insights into the behaviour and design of AOGCM hosing experiments. However as a caveat we note that recently Haskins et al. \cite{Haskins18} have analysed a ``press" scenario in the FAMOUS AOGCM, in a case which shows an oscillatory AMOC recovery after the forcing is removed. While this oscillatory recovery looks qualitatively similar to our box model solutions, it appears to be driven by variations in the South Atlantic density, a processes that is not captured in the three box, and possibly not the five box model. Other processes may operate in AOGCMs that produce qualitatively similar looking dynamics to the box model solutions. 

Finally we discuss the rate-dependent tipping observed in Section~\ref{sec:Rtip}. 
The concept of rate-dependent tipping in response to fresh water hosing was explored in a two-dimensional Boussinesq model in \cite{LucariniCalmanti2005,LucariniCalmanti2007}, where it was shown that fast perturbations can destabilize the typical advective (negative) feedback of the AMOC. 
Additionally, Stocker and Schmittner \cite{stocker1997} found rate-dependent AMOC tipping in an intermediate-complexity climate model (EMIC) in response to increasing greenhouse gases. Faster $\mathrm{CO}_2$ increase could result in tipping, while a more gradual increase toward the same steady state value did not. Although the $\mathrm{CO}_2$ forcing used by \cite{stocker1997} is not the same as the fresh water forcing we use, we note from Figure~\ref{fig:tip_before_hom} that similar behaviour can be seen in the box model, supporting the view that in order to predict the future stability of the AMOC we need to understand not just the sustainability of an equilibrium ``on" state under climate forcing, but also the rate of forcing change (i.e. the emissions pathway, not just the final greenhouse gas levels).

\section{Discussion}
\label{sec:discussion}

In this paper, we investigate the global ocean box model of \cite{RWSH17} and demonstrate, similarly to \cite{Titzetal2002a}, that the switching from the ``on" state is associated with subcritical Hopf rather than saddle-node bifurcation. We interpret time-dependent perturbations in terms of the bifurcation diagrams and show in particular that there can be a mixing of rate- and bifurcation-tipping.

While the forcing scenarios we investigate are highly idealised, they illustrate the possibility of rate-dependent tipping in the climate system, which is of particular interest as the international community considers options to limit dangerous climate change under the Paris climate accord. Our results reveal that tipping can occur in response to a rapid increase in forcing, even if slower increase to the same level of forcing does not result in tipping. Conversely if a forcing threshold is passed such that the AMOC is ultimately unsustainable, tipping may still be avoided if the forcing increase is reversed quickly enough. These results suggest a framework for the design of safe climate stabilisation pathways to avoid AMOC tipping. The characteristic direction and curvature of the phase trajectories in $(S_N,S_T)$ space is potentially observable in the real ocean and so could provide early warning that the AMOC is on the attractor of an ``off'' state.

Our work needs to be followed up using more realistic scenarios and models, but the theoretical understanding developed here can be used to guide experimentation with more comprehensive (but computationally expensive) models to address the questions "Under what circumstances might the AMOC switch to an off state?", and "What would we need to observe to give the earliest possible warning of such a switch?" 

We highlight that because the bifurcation to a finite basin  occurs before loss of stability, this means that even perturbations that are not large enough to cause tipping may result in significant and long deviations, even if there is eventually a return. For example, in Figure~\ref{fig:threebox_basins_H}~(c) if the state of the system is perturbed such that it gets close enough to the boundary of the basin of $X_\mathrm{on}$, it needs more than 2000 years to get back to equilibrium. For this reason it seems plausible that the present day AMOC may still retain significant memory of major perturbations that it underwent hundreds of years ago. 

For our investigations we only consider time-dependent perturbation of the ``hosing" parameter $H$ where it is assumed that fluxes are balanced and so total ocean volume and salt content are preserved. Over longer timescales or for larger perturbations ({\it{e.g.}} glacial-interglacial transitions) this approximation will not necessarily be valid. In particular (\ref{eq:totalsalt}) will no longer be constant and the associated dimension reduction not valid. Potentially this will give rise to new dynamical effects.

Finally we highlight the  close relationship between R-tipping and nonautonomous stability: \cite{Alkhayuon2017,Ashwin2017} show that R-tipping can be understood as a change in properties of a local pullback attractor and identify criteria for the appearance of R-tipping in simple systems. R-tipping can also appear in the presence of noise, for example \cite{Ritchie2016,Ritchie2017} have studied the probability of the occurrence of R-tipping in the presence of noise, as well as some potential early warning indicators of such a phenomenon. Other recent work has examined in detail the crossing ``points of no return'' for systems with R-tipping \cite{OKeefe2019,Ritchie2019}, but in general the science of how reliable early warning signals may be for R-tipping is still to be worked out.


\enlargethispage{20pt}





\subsection*{Acknowledgements}

The simulation code used to generate figures for the box models, and the movies, are available via the code repository {\tt https://github.com/peterashwin/AAJQW\_AMOC}.

HA was supported by the Higher Committee For Education Development in Iraq (HCED Iraq) via grant agreement No D13436.  CQ was supported by the CRITICS Innovative training Network for funding via the European Union's Horizon 2020 research and innovation programme under the Marie Sk{\l}odowska-Curie grant agreement No 643073. LJ and RW were supported by the Met Office Hadley Centre Climate Programme funded by BEIS and Defra. All authors thank the Past Earth Network (EPSRC grant number EP/M008363/1) and ReCoVER (EPSRC grant number EP/M008495/1) for supporting a workshop in 2017 where this work started.

Thanks also to Ulrike Feudel, Felix Mendelssohn, Paul Ritchie, Jan Sieber, Sebastian Wieczorek for inspiration and insightful comments.



\vskip2pc

\bibliographystyle{RS}






\begin{thebibliography}{99}

\bibitem{Stommel1961}
Stommel H. 1961  Thermohaline Convection with Two Stable Regimes of Flow. {\em
  Tellus} \textbf{13}, 224--230.

\bibitem{Dijkstra2007}
Dijkstra HA. 2007  Characterization of the multiple equilibria regime in a
  global ocean model. {\em Tellus A} \textbf{59}, 695--705.

\bibitem{Dijkstra2013}
Dijkstra HA. 2013 {\em Nonlinear climate dynamics}.
Cambridge University Press, Cambridge.

\bibitem{hawkins2011}
Hawkins E, Smith RS, Allison LC, Gregory JM, Woollings TJ, Pohlmann H,
  De~Cuevas B. 2011  Bistability of the Atlantic overturning circulation in a
  global climate model and links to ocean freshwater transport. {\em
  Geophysical Research Letters} \textbf{38}.

\bibitem{Rahmstorf1996}
Rahmstorf S. 1996  On the freshwater forcing and transport of the Atlantic
  thermohaline circulation. {\em Climate Dynamics} \textbf{12}, 799--811.

\bibitem{Rahmstorf1999}
Rahmstorf S. 1999  Rapid transitions of the thermohaline ocean circulation. In
  {\em Reconstructing Ocean History} pp. 139--149. Springer.

\bibitem{rahmstorf2005thermohaline}
Rahmstorf S, Crucifix M, Ganopolski A, Goosse H, Kamenkovich I, Knutti R,
  Lohmann G, Marsh R, Mysak LA, Wang Z, Weaver AJ. 2005  Thermohaline
  circulation hysteresis: A model intercomparison. {\em Geophysical Research
  Letters} \textbf{32}.

\bibitem{Bondetal:2013}
Bond GC, Showers W, Elliot M, Evans M, Lotti R, Hajdas I, Bonani G, Johnson S.
  2013 pp. 35--58.
In {\em The North Atlantic's 1-2 Kyr Climate Rhythm: Relation to Heinrich
  Events, Dansgaard/Oeschger Cycles and the Little Ice Age}, pp. 35--58.
  American Geophysical Union (AGU).

\bibitem{ClementPeterson2008}
Clement AC, Peterson LC. 2008  Mechanisms of abrupt climate change of the last
  glacial period. {\em Reviews of Geophysics} \textbf{46}.

\bibitem{Scheffer2009}
Scheffer M, Bascompte J, Brock WA, Brovkin V, Carpenter SR, Dakos V, Held H,
  van Nes EH, Rietkerk M, Sugihara G. 2009  {Early-warning signals for critical
  transitions}. {\em Nature} \textbf{461}, 53--59.

\bibitem{Lenton2011}
Lenton TM. 2011  {Early warning of climate tipping points}. {\em Nature Climate
  Change} \textbf{1}, 201--209.

\bibitem{Ashwin2011}
Ashwin P, Wieczorek S, Vitolo R, Cox P. 2012  {Tipping points in open systems:
  bifurcation, noise-induced and rate-dependent examples in the climate
  system}. {\em Phil. Trans. R. Soc. A} \textbf{370}, 1166--1184.

\bibitem{Scheffer2008}
Scheffer M, {Van Nes} EH, Holmgren M, Hughes T. 2008  {Pulse-driven loss of
  top-down control: The critical-rate hypothesis}. {\em Ecosystems}
  \textbf{11}, 226--237.

\bibitem{vbB2018}
Bolt, B and Nes, EH and Bathiany, S and Vollebregt, ME and Scheffer, M. 2018
Climate reddening increases the chance of critical transitions, {\em Nature Climate Change}, {\bf 8}:478.

\bibitem{JW17}
Jackson L, Wood R. 2018  Hysteresis and Resilience of the AMOC in an
  Eddy-Permitting GCM. {\em Geophysical Research Letters} \textbf{45},
  8547--8556.

\bibitem{stocker1997}
Stocker T, Schmittner A. 1997  Influence of $CO_2$ emission rates on the
  stability of the thermohaline circulation. {\em Nature} \textbf{388},
  862--865.

\bibitem{mccarthy2015measuring}
McCarthy GD, Smeed DA, Johns WE, Frajka-Williams E, Moat BI, Rayner D, Baringer
  MO, Meinen CS, Collins J, Bryden HL. 2015  {Measuring the Atlantic Meridional
  Overturning Circulation at 26$^o$N}. {\em Progress in Oceanography}
  \textbf{130}, 91--111.

\bibitem{RWSH17}
Wood R, Rodriguez J, Smith R, Jackson L, Hawkins E. 2019  Observable, low-order
  dtnamical controls on thresholds of the Atlantic Meridional Overturning
  Circulation. {\em Climate Dynamics (submitted)}.

\bibitem{Peltier:2014}
Richard PW, Guido V. 2014  Dansgaard-Oeschger oscillations predicted in a
  comprehensive model of glacial climate: A 'kicked' salt oscillator in the
  Atlantic. {\em Geophysical Research Letters} \textbf{41}, 7306--7313.

\bibitem{Lohmann1999}
Lohmann G, Schneider J. 1999  Dynamics and predictability of Stommel's box
  model. A phase-space perspective with implications for decadal climate
  variability. {\em Tellus A: Dynamic Meteorology and Oceanography}
  \textbf{51}, 326--336.

\bibitem{LucariniStone2005}
Lucarini V, Stone PH. 2005  Thermohaline Circulation Stability: A Box Model
  Study. Part I: Uncoupled Model. {\em Journal of Climate} \textbf{18},
  501--513.
  
 \bibitem{LucariniCalmanti2005}
Lucarini V, Calmanti S, Artale V. 2005  
Destabilization of the thermohaline circulation by transient changes in the hydrological cycle. {\em Climate Dynamics} \textbf{24},
  253-262.
  
\bibitem{LucariniCalmanti2007} 
Lucarini V, Calmanti S, Artale V. 2007  
Experimental Mathematics: Dependence
of the Stability Properties of a Two-Dimensional Model of the Atlantic Ocean
Circulation on the Boundary Conditions {\em Russ. J. Math. Phys} \textbf{14},
 224--231.
 
\bibitem{LucariniFaranda2012}
Lucarini V, Faranda D, Willeit M. 2012  Bistable systems with stochastic noise: virtues and limits of effective one-dimensional Langevin equations {\em Nonlin. Processes Geophys.} \textbf{19},
 9--22.
 
\bibitem{Titzetal2002a}
Titz S, Kuhlbrodt T, Rahmstorf S, Feudel U. 2002a  On freshwater-dependent
  bifurcations in box models of the interhemispheric thermohaline circulation.
  {\em Tellus A} \textbf{54}, 89--98.

\bibitem{Titzetal2002b}
Titz S, Kuhlbrodt T, Feudel U. 2002b  Homoclinic bifurcation in an ocean
  circulation box model. {\em International Journal of Bifurcation and Chaos}
  \textbf{12}, 869--875.

\bibitem{jacksonfamous17}
Jackson L, Smith R, Wood R. 2017  Ocean and atmosphere feedbacks affecting AMOC
  hysteresis in a GCM. {\em Climate Dynamics} \textbf{49}, 173--191.

\bibitem{DeVries05}
de~Vries P, Weber SL. 2005  {The Atlantic freshwater budget as a diagnostic for
  the existence of a stable shut down of the Meridional Overturning
  Circulation}. {\em Geophys. Res. Lett} \textbf{32}.

\bibitem{liu2012diagnostic}
Liu W, Liu Z. 2012  {A Diagnostic Indicator of the Stability of the Atlantic
  Meridional Overturning Circulation in CCSM3}. {\em J. Climate} \textbf{26},
  1926--1938.

\bibitem{manabe88}
Manabe S, Stouffer R. 1988  Two stable equilibria of a coupled ocean-atmosphere
  model. {\em J. Climate} \textbf{1}, 841--863.

\bibitem{jackson18ClDyn}
Jackson L, Wood R. 2018  Timescales of AMOC decline in response to fresh water
  forcing. {\em Climate Dyn.} \textbf{51}, 1333--1350.

\bibitem{sijp2012precise}
Sijp WP, England MH. 2012  Precise Calculations of the Existence of Multiple
  AMOC Equilibria in Coupled Climate Models. Part II: Transient Behavior. {\em
  Journal of Climate} \textbf{25}, 299--306.

\bibitem{Smith2012}
Smith RS. 2012  The FAMOUS climate model (versions XFXWB and XFHCC):
  description update to version XDBUA. {\em Geoscientific Model Development}
  \textbf{5}, 269--276.

\bibitem{Dumortier2000}
Dumortier F, Herssens C, Perko L. 2000  Local Bifurcations and a Survey of
  Bounded Quadratic Systems. {\em Journal of Differential Equations}
  \textbf{165}, 430 -- 467.

\bibitem{Bernardo2008}
Bernardo M, Budd C, Champneys AR, Kowalczyk P. 2008 {\em Piecewise-smooth
  dynamical systems: theory and applications} vol. 163.
Springer Science \& Business Media.

\bibitem{Kowalczyk2011}
Kowalczyk P, Glendinning P. 2011  Boundary-equilibrium bifurcations in
  piecewise-smooth slow-fast systems. {\em Chaos: An Interdisciplinary Journal
  of Nonlinear Science} \textbf{21}, 023126.

\bibitem{cessi1994simple}
Cessi, P. 1994 
A simple box model of stochastically forced thermohaline flow. {\em Journal of physical oceanography} \textbf{24}, 1911--1920.

\bibitem{xppautref}
Ermentrout B. 2002 {\em Simulating, analyzing, and animating dynamical systems}
  vol.~14{\em Software, Environments, and Tools}.
Society for Industrial and Applied Mathematics (SIAM), Philadelphia, PA.
A guide to XPPAUT for researchers and students.

\bibitem{cocoref}
Dankowicz H, Schilder F. 2013 {\em Recipes for continuation} vol.~11{\em
  Computational Science \& Engineering}.
Society for Industrial and Applied Mathematics (SIAM), Philadelphia, PA.

\bibitem{Kuznetsov1998}
Kuznetsov YA. 1998 {\em Elements of Applied Bifurcation Theory (2nd Ed.)}.
Berlin, Heidelberg: Springer-Verlag.

\bibitem{Kuehn2015}
Kuehn C. 2015 {\em {Multiple Time Scale Dynamics}}.
Springer, New York.

\bibitem{Ritchie2019}
Ritchie P, Karabacak \"{O}, Sieber J. 2019  {Inverse-square law between time and amplitude for crossing tipping thresholds}.
{\em Proceedings of the Royal Society A} \textbf{475}, 20180504.

\bibitem{ratajczak2017}
Ratajczak Z, D'odorico P, Collins SL, Bestelmeyer BT, Isbell FI, Nippert JB.
  2017  The interactive effects of press/pulse intensity and duration on regime
  shifts at multiple scales. {\em Ecological Monographs} \textbf{87}, 198--218.

\bibitem{Haskins18}
Haskins R, Oliver K, Jackson L, Drijfhout S, Wood R. 2018  Explaining asymmetry
  between weakening and recovery of the AMOC in a coupled climate model. {\em
  Climate Dyn.}

\bibitem{Alkhayuon2017}
Alkhayuon HM, Ashwin P. 2018  Rate-induced tipping from periodic attractors:
  Partial tipping and connecting orbits. {\em Chaos} \textbf{28}, 033608.

\bibitem{Ashwin2017}
Ashwin P, Perryman C, Wieczorek S. 2017  {Parameter shifts for nonautonomous
  systems in low dimension: Bifurcation- and Rate-induced tipping}. {\em
  Nonlinearity} \textbf{30}, 2185--2210.

\bibitem{Ritchie2016}
Ritchie P, Sieber J. 2016  {Early-warning indicators for rate-induced tipping}.
  {\em Chaos} \textbf{26}.

\bibitem{Ritchie2017}
Ritchie P, Sieber J. 2017  {Probability of noise-and rate-induced tipping}.
  {\em Physical Review E} \textbf{95}, 1--13.

\bibitem{OKeefe2019}
O'Keefe PE, Wieczorek S. 2019 {Tipping Phenomena and Points of No Return in Ecosystems: Beyond Classical Bifurcations}.
  {\em arXiv}:1902.01796.


\end{thebibliography}

\appendix

\section{Model constants and parameters \label{sec:appdx1}}

The default constants and parameters for the five box model are listed in Tables~\ref{tab:oceans} and \ref{tab:params}.

\begin{table}
$$
\begin{array}{l|rrr}
 & \mbox{Volume} & \mbox{Salinity}& \mbox{Flux} \\
 \hline
\mbox{North Atlantic} & V_N= 0.3261 \x 10^{17} \mcubed & S_N = 0.034912 & F_N= 0.384 \Sv \\
\mbox{Tropical Atlantic} & V_T= 0.7777 \x 10^{17}  \mcubed  & S_T = 0.035435 & F_T= -0.723 \Sv \\
\mbox{Southern Ocean} & V_S= 0.8897 \x 10^{17}  \mcubed  & S_S = 0.034427 & F_S= 1.078 \Sv \\
\mbox{Indo-Pacific} & V_{IP}= 2.2020 \x 10^{17}  \mcubed  & S_{IP} = 0.034668 & F_{IP}=-0.738 \Sv \\
\mbox{Bottom Ocean} & V_{B}= 8.6490 \x 10^{17}  \mcubed  & S_{B} = 0.034538 &
\end{array}
$$
\caption{The volumes and standard (baseline, $1\times \mathrm{CO}_2$) values for the AMOC models, based on the FAMOUS$_{B}$ runs \cite{Smith2012} that are used in \cite{RWSH17}. Note that for the five box model all Salinities vary with time so these are initial conditions while for the three box model the Salinities $S_T,S_N$ and $S_B$ vary with time; the others are fixed parameters. Note that the fluxes are assumed to balance: $F_N+F_T+F_S+F_{IP}=0$. }
\label{tab:oceans}
\end{table}

\begin{table}
$$
\begin{array}{l|rlc||l|rl}
\mbox{Name} & \mbox{Default value} & \mbox{Units}&~~&\mbox{Name} & \mbox{Default value} & \mbox{Units}\\
\hline
\alpha & 0.12 &\kg \mneg \degC^{-1}&&K_N & 5.456 &\Sv  \\
\beta & 790.0 &\kg \mneg &&K_S & 5.447 &\Sv \\
S_0 & 0.035& && K_{IP} & 96.817 &\Sv \\
T_S & 4.773 &\degC && \lambda & 2.79 \x 10^{7}  &\mathrm{m}^{6} \kg^{-1} \mathrm{s}^{-1} \\
T_0 & 2.650 &\degC &&\gamma & 0.39&\\
\eta & 74.492 &\Sv && \mu & 5.5 &  \degC ^{-1} \mathrm{m}^{-3} \mathrm{s}\x10^{-8}
\end{array}
$$
\caption{Baseline parameters  ($1\times \mathrm{CO}_2$) used in the model, taken from FAMOUS$_{B}$ runs \cite{Smith2012} in \cite{RWSH17}. Note that $1 \Sv= 10^6 \mathrm{m}^3 \mathrm{s}^{-1}$.}
\label{tab:params}
\end{table}

\begin{table}
$$
\begin{array}{l|rrr}
 & \mbox{Volume} & \mbox{Salinity}& \mbox{Flux} \\
 \hline
\mbox{North Atlantic} & V_N= 0.3683 \x 10^{17} \mcubed & S_N = 0.034912 & F_N= 0.486 \Sv \\
\mbox{Tropical Atlantic} & V_T= 0.5418 \x 10^{17}  \mcubed  & S_T = 0.035435 & F_T= -0.997 \Sv \\
\mbox{Southern Ocean} & V_S= 0.6097 \x 10^{17}  \mcubed  & S_S = 0.034427 & F_S= 1.265 \Sv \\
\mbox{Indo-Pacific} & V_{IP}= 1.4860 \x 10^{17}  \mcubed  & S_{IP} = 0.034668 & F_{IP}=-0.754 \Sv \\
\mbox{Bottom Ocean} & V_{B}= 9.9250 \x 10^{17}  \mcubed  & S_{B} = 0.034538 &
\end{array}
$$
\caption{The volumes and standard (baseline) values for the AMOC models with doubled atmospheric $\mathrm{CO}_2$ ($2\times \mathrm{CO}_2$), based on the FAMOUS$_{B}$ runs \cite{Smith2012} that are used in \cite{RWSH17}. Note that for the five box model all Salinities vary with time so these are initial conditions while for the three box model the Salinities $S_T,S_N$ and $S_B$ vary with time; the others are fixed parameters. Note that the fluxes are assumed to balance: $F_N+F_T+F_S+F_{IP}=0$. }
\label{tab:oceans2CO2}
\end{table}

\begin{table}
$$
\begin{array}{l|rlc||l|rl}
\mbox{Name} & \mbox{Default value} & \mbox{Units}&~~&\mbox{Name} & \mbox{Default value} & \mbox{Units}\\
\hline
\alpha & 0.12 &\kg \mneg \degC^{-1}&&K_N & 1.762 &\Sv  \\
\beta & 790.0 &\kg \mneg &&K_S & 1.872 &\Sv \\
S_0 & 0.035& && K_{IP} & 99.977 &\Sv \\
T_S & 7.919 &\degC && \lambda & 1.62 \x 10^{7}  &\mathrm{m}^{6} \kg^{-1} \mathrm{s}^{-1} \\
T_0 & 3.870 &\degC &&\gamma & 0.36&\\
\eta & 33.264 &\Sv && \mu & 22 &  \degC ^{-1} \mathrm{m}^{-3} \mathrm{s}\x10^{-8}
\end{array}
$$
\caption{Parameters used in the model with doubled atmospheric $\mathrm{CO}_2$ ($2\times \mathrm{CO}_2$), taken from FAMOUS$_{B}$ runs \cite{Smith2012} in \cite{RWSH17}. Note that $1 \Sv= 10^6 \mathrm{m}^3 \mathrm{s}^{-1}$.}
\label{tab:params2CO2}
\end{table}

\section{The scaled five box model \label{sec:appdx2}}

In order to improve computational efficiency (particularly for the continuation software), we rescale both time and state variables of the models.  The time unit $t$ of the original model is in seconds; we will consider a time unit of years $\tau = tY^{-1}$ where $Y=3.15\x 10^{7}$.  In addition we will consider the state variables as scaled perturbations from a background state, namely 
\begin{equation}
\label{eq:scaled_S}
\tilde{S}_i = 100(S_i - S_0), \qquad i\in[N,T,S,IP,B].
\end{equation}
This leads to a modified equation for $q$,
$$
q=\lambda [\alpha (T_S - T_N) + \frac{\beta}{100} (\tilde{S}_N - \tilde{S}_S)],
$$
and the following ODEs:
For $q \geq 0$ we have
\begin{equation}
\label{eq:AOMC_model_scaled_q>0}
\left.\begin{array}{rcl}
\frac{V_N}{Y} \frac{d\tilde{S}_N}{d\tau}	&=& q(\tilde{S}_T - \tilde{S}_N) + K_N (\tilde{S}_T - \tilde{S}_N) - 100 F_N S_0, \\
\frac{V_T}{Y} \frac{d\tilde{S}_T}{d\tau}	&=& q[\gamma \tilde{S}_S + (1- \gamma) \tilde{S}_{IP} - \tilde{S}_T] + K_S (\tilde{S}_S - \tilde{S}_T)+K_N(\tilde{S}_N - \tilde{S}_T) - 100 F_T S_0,\\
\frac{V_S}{Y} \frac{d\tilde{S}_S}{d\tau}	&=& \gamma q (\tilde{S}_B -\tilde{S}_S) + K_{IP} (\tilde{S}_{IP} - \tilde{S}_S) + K_S (\tilde{S}_T - \tilde{S}_S) +\eta (\tilde{S}_B - \tilde{S}_S) - 100 F_S S_0, \\
\frac{V_{IP}}{Y} \frac{d\tilde{S}_{IP}}{d\tau} &=& (1-\gamma)q(\tilde{S}_B - \tilde{S}_{IP}) + K_{IP} (\tilde{S}_S - \tilde{S}_{IP}) - 100 F_{IP} S_0, \\
\frac{V_{B}}{Y} \frac{d\tilde{S}_B}{d\tau} &=& q (\tilde{S}_N - \tilde{S}_B) + \eta (\tilde{S}_S - \tilde{S}_B).
\end{array}\right\}
\end{equation}
while for $q < 0$ we have
\begin{equation}
\label{eq:AOMC_model_scaled_q<0}
\left.\begin{array}{rcl}
\frac{V_N}{Y} \frac{d\tilde{S}_N}{d\tau}	 &=& |q|(\tilde{S}_B - \tilde{S}_N) + K_N (\tilde{S}_T - \tilde{S}_N) - 100 F_N S_0, \\
\frac{V_T}{Y} \frac{d\tilde{S}_T}{d\tau}	&=& |q| (\tilde{S}_N - \tilde{S}_T) + K_S (\tilde{S}_S - \tilde{S}_T)+K_N(\tilde{S}_N - \tilde{S}_T) - 100 F_T S_0,\\
\frac{V_S}{Y} \frac{d\tilde{S}_S}{d\tau}	&=& \gamma |q| (\tilde{S}_T -\tilde{S}_S) + K_{IP} (\tilde{S}_{IP} - \tilde{S}_S) + K_S (\tilde{S}_T - \tilde{S}_S) +\eta (\tilde{S}_B - \tilde{S}_S) - 100 F_S S_0, \\
\frac{V_{IP}}{Y} \frac{d\tilde{S}_{IP}}{d\tau} &=& (1-\gamma)|q|(\tilde{S}_T - \tilde{S}_{IP}) + K_{IP} (\tilde{S}_S - \tilde{S}_{IP}) - 100 F_{IP} S_0, \\
\frac{V_{B}}{Y} \frac{d\tilde{S}_B}{d\tau} &=& \gamma |q| \tilde{S}_S + (1-\gamma) |q| \tilde{S}_{IP} - |q| \tilde{S}_B + \eta (\tilde{S}_S -\tilde{S}_B).
\end{array}\right\}
\end{equation}
Note that the total salt content is still conserved with
\begin{equation}
C= \frac{1}{100}(V_N \tilde{S}_N+ V_T \tilde{S}_T+ V_S \tilde{S}_S+ V_{IP} \tilde{S}_{IP} + V_B \tilde{S}_B)+S_0(V_N+V_T+V_S+V_{IP}+V_B).
\label{eq:totalsalt_scaled}
\end{equation}
We have only shown the scaled version of the five box model but the three box model is scaled equally.

\section{Supplementary Movies}

Supplementary Movie {\tt MOC\_basin\_2co2.mp4} shows animations of Figure~\ref{fig:threebox_basins_H}. Supplementary Movies {\tt MOC\_phas\_Btip.mp4} and {\tt MOC\_phas\_Btrack.mp4} show animations corresponding to the red and black trajectories respectively for Figure~\ref{fig:Btip}. Supplementary Movies {\tt MOC\_phas\_Rtip.mp4} and {\tt MOC\_phas\_Rtrack.mp4} show animations corresponding to the red and black trajectories respectively for Figure~\ref{fig:Rtip}.

\end{document}